\documentclass{article}

\PassOptionsToPackage{numbers,sort&compress}{natbib}
 \usepackage[preprint]{neurips_2026}

\usepackage[utf8]{inputenc} 
\usepackage[T1]{fontenc}    
\usepackage[hidelinks]{hyperref}       
\usepackage{url}            
\usepackage{booktabs}       
\usepackage{amsfonts}       
\usepackage{nicefrac}       
\usepackage{microtype}      
\usepackage{xcolor}         
\usepackage{amsmath}

\usepackage{enumitem}

\usepackage{lipsum}

\usepackage{todonotes}
\usepackage{subcaption}
\usepackage{graphicx}
\usepackage{caption}
\usepackage{booktabs, longtable,array, tabularx, makecell}
\usepackage{tcolorbox}
\tcbuselibrary{breakable, skins}
\newtcolorbox{promptbox}[1][]{
  breakable,
  enhanced,
  colback=gray!8,
  colframe=gray!50,
  fonttitle=\small\bfseries,
  title={#1},
  left=6pt, right=6pt, top=4pt, bottom=4pt,
  boxrule=0.5pt,
}
\usepackage{listings}
\title{TSAgent: An Agentic Workflow for Autonomous Transition State Search}

\author{%
  Varun Madhavan \\
  Department of Chemical Engineering \\
  University of Michigan Ann Arbor \\
  Ann Arbor, MI 48109 \\
  \texttt{varunvm@umich.edu} \\
  \And
  Ankit Mathanker \\
  Department of Chemical Engineering \\
  University of Michigan Ann Arbor \\
  Ann Arbor, MI 48109 \\
  \texttt{ankitma@umich.edu} \\
  \And
  Dean M. Sweeney \\
  Department of Chemical Engineering \\
  University of Michigan Ann Arbor \\
  Ann Arbor, MI 48109 \\
  \texttt{dmsween@umich.edu} \\
  \And
  Oluwatosin A. Ohiro \\
  Department of Chemical Engineering \\
  University of Michigan Ann Arbor \\
  Ann Arbor, MI 48109 \\
  \texttt{oohiro@umich.edu} \\
  \And
  Yixin Wang \\
  Department of Statistics \\
  University of Michigan Ann Arbor \\
  Ann Arbor, MI 48109 \\
  \texttt{yixinw@umich.edu} \\
  \And
  Bryan R. Goldsmith \\
  Department of Chemical Engineering \\
  University of Michigan Ann Arbor \\
  Ann Arbor, MI 48109 \\
  \texttt{bgoldsm@umich.edu} \\
}

\begin{document}

\maketitle

\begin{abstract}

   Identifying transition states (TSs) on potential energy surfaces is a central computational bottleneck in mechanistic studies of catalytic materials. A TS search is not a single calculation but a long-horizon, multi-step workflow of atomistic simulations with delayed, asynchronous feedback and heterogeneous failure modes that require a joint multimodal analysis of scalar convergence diagnostics and atomic geometries along the reaction path. To address this challenge, we propose \textbf{TSAgent}, an agentic workflow that automates TS search directly at the density functional theory (DFT) level of quantum chemical accuracy. TSAgent operates through a persistent plan–execute–analyze–replan loop, continuously adapting its strategy based on convergence diagnostics and geometric feedback without human intervention. We evaluate TSAgent on a diverse 100-example subset of the OC20NEB heterogeneous catalysis benchmark, where it successfully locates TSs with 83\% accuracy. In a direct comparison against expert DFT practitioners on 10 held-out examples, TSAgent achieves a 70\% success rate compared to a human-expert average of $73 \pm 12$\%. Finally, TSAgent independently reproduces Brønsted–Evans–Polanyi scaling relationships for NH$_3$ dissociation on metal and single-atom alloy surfaces from a published heterogeneous catalysis study, demonstrating that its utility extends beyond curated benchmarks to real scientific investigations.


\end{abstract}

\section{Introduction}

\begin{figure}[t]
  \centering 
  \includegraphics[width=1\linewidth]{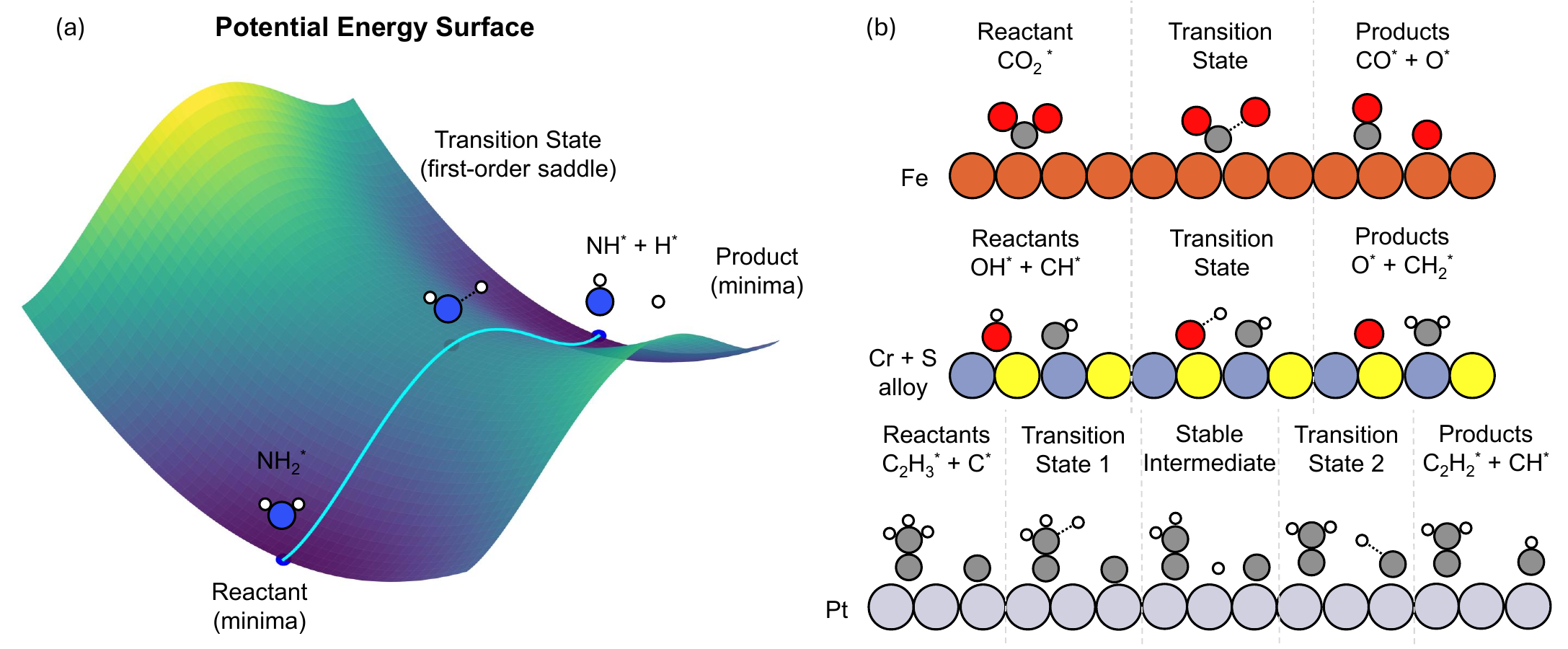}
  \caption{(\textit{a}) A schematic potential energy surface (PES), showing a dissociation reaction with $NH_2^*$ as the reactant, $NH \cdots H^*$ as the transition state (bond breaking), and $NH^* + H^*$ as the product state. Stable configurations (reactants/products) correspond to local minima, while transition states are first-order saddle points that separate them. The activation barrier, that is, the energy difference between the reactant minimum ($NH_2^*$) and the adjacent saddle point ($NH \cdots H^*$), controls the reaction rate.  Here $*$ denotes a species adsorbed to a surface. (\textit{b}) Three representative heterogeneous catalysis reactions on different metal surfaces, showing the atomistic configurations at the reactant, transition state, and product geometries. Reaction 1 exhibits $CO_2^*$ dissociation to form $CO^*$ and $O^*$ (i.e., a dissociation reaction). Reaction 2 is a transfer reaction, where $OH^*$ donates its $H^*$ directly to $CH^*$ on the surface. Reaction 3 defines a multi-step reaction, where $C_2H_3^*$ first dissociates to form $C_2H_2^*$ and $H^*$, which then forms $CH^*$ with an adsorbed carbon atom, $C^*$.}
  \label{fig:pes}
      \vspace{-6pt}
\end{figure}

Chemical reactions arise from molecular motion on potential energy surfaces (PES), which are $3N-6$ dimensional functions of the nuclear coordinates of an $N$-atom system that describe stable states and energy barriers between molecules, as shown in Figure~\ref{fig:pes}a. Understanding the mechanisms underlying chemical reactions is a crucial step in the rational design of catalysts, enzymes, and functional materials \cite{simm-auto-exp2022, broderick-deep-exp2022}. Computational mechanistic studies probe the PES by identifying key stationary points, including local minima corresponding to stable intermediates and first-order saddle points corresponding to transition states (TSs) \cite{schlegel_geometry_2011}. The TSs are the highest energy points along minimum energy paths connecting two minima; they define the activation barriers for elementary steps (e.g. individual mechanistic events between two minima like bond breaking) that govern reaction rates and product selectivity. 

Computational mechanistic studies widely use Density Functional Theory (DFT) simulations to study the PES and find TSs \cite{dft-review}. Since DFT is a first-principles method that derives system energies and atomic forces via solutions to the Kohn–Sham equations \cite{KohnSham1965}, each calculation is computationally intensive with runtimes ranging from hours to days on massively parallel high-performance computing (HPC) hardware \cite{ocp-neb}. Mechanistic studies require a sequence of these computationally intensive DFT calculations to produce a detailed atomistic picture of how and why a reaction proceeds.

 Locating the TS of reactions on heterogeneous catalysts is among the most manually intensive tasks in computational chemistry \cite{cheula_fine-tuning_2026}. A TS search is not a single DFT calculation but a multi-stage optimization pipeline, where each stage is susceptible to qualitatively distinct failure modes \cite{Greten-Benchmarking2026}. Critically, many failure modes are not detectable solely by scalar convergence criteria and require a joint analysis of multiple DFT output metrics, such as system energies, forces, and vibrational frequencies, as well as direct \textit{visual inspection} of the 3D atomic configurations along the reaction path. Because each failure mode demands a distinct, physics-informed corrective intervention, no single automated recovery strategy is effective at finding TSs at scale. Furthermore, to effectively explore the catalyst design space, mechanistic studies must account for the combinatorial explosion of TSs between competing elementary steps \cite{Morandi2026CARE}. For example, the catalytic reduction of CO$_2$ and nitrogenous species (N$_2$, NO$_3^-$, NO$_2^-$) into urea (CO(NH$_2$)$_2$) involves a large number of adsorbed C- and N-containing intermediates along each individual reduction pathway \cite{Li2024UreaElectrosynthesis}. These intermediates can combine on the catalyst surface through numerous C--N bond-forming events (e.g., $CO^*$ + $N_2^*$, $CO^*$ + $NH_x^*$, $CHO^*$ + $N_2^*$, $CO^*$ + $N^*$, $COH^*$ + $N_2^*$, etc.) to form secondary intermediates (e.g., $OCNO^*$, $NCON^*$, $NCO^*$,  etc.) \cite{li_photochemical_2026}. 

The difficulty of finding TSs, compounded by the combinatorial explosion of candidate TSs, makes it the central bottleneck for large-scale mechanistic studies and, by extension, catalyst and materials design. The manually intensive nature of TS search workflows makes it infeasible to automate via static scripts. Instead, we need a system capable of reasoning through failure---interpreting multimodal simulation outputs, diagnosing the actual cause of a failure, and adapting the strategy in response. Agentic workflows, with the capability to reason over domain knowledge, multimodal simulation outputs, and instructions from experts, are a promising framework for automating TS searches. In this work, we present an agentic workflow to automate the full suite of DFT calculations for identifying TSs in heterogeneous catalysis reactions, with the capability to autonomously diagnose and recover from failures as required to find a theoretically validated TS geometry. We summarize our contributions as follows:
\begin{itemize}[leftmargin=0.5cm, topsep=-0.05cm]
    \item We introduce TSAgent, an agentic workflow for TS searching with a closed-loop, multimodal reasoning process that mirrors how human practitioners diagnose failed TS search calculations. The agent combines diagnostics from DFT outputs with visual analyses of reaction events to identify physically meaningful events such as bond formation and breaking, rotation, desorption, atomic collisions, and intermediate stabilization, and autonomously revises its strategy during failure events to find physically validated TS geometries. 
    
    \item We demonstrate that the workflow autonomously identifies TS geometries with an accuracy of 83\% across a sample of 100 reaction pathways from the OC20NEB benchmark dataset \cite{ocp-neb}, spanning multiple transition metals, surface facets, and reaction types. Unlike prior automated TS pipelines that leverage ML-based approximations \cite{ocp-neb, nebscape}, our system closes the loop directly around DFT, identifying TSs at the same level of theory used to construct the benchmark.

    \item We benchmark the agent against three expert DFT practitioners, demonstrating that the agent matches expert-level success rates (70\% vs. $73 \pm 12$\%) without any human intervention. To our knowledge, this is the first quantitative comparison of an autonomous agent against domain experts on the TS search task.

    \item Finally, we reproduce the Br{\o}nsted--Evans--Polanyi scaling relations for NH$_3$ dissociation on metal and single-atom alloy surfaces from a published study \cite{darby_elucidating_2018}, demonstrating that TSAgent can autonomously execute real-world TS searches of scientific value.
    
\end{itemize}

\footnotetext{The full implementation of TSAgent, including the agent code, prompts, and workflow configurations is available at \url{https://github.com/transition-state-search/TSAgent}.}

\section{Related Work}
\label{section:related-work}


Several works have demonstrated the agentic orchestration of DFT calculations. DREAMS \cite{dreams} introduced a hierarchical multi-agent framework for DFT-based materials simulations including lattice constant predictions and surface adsorption calculations, comprising a central planner and domain-specific sub-agents for structure generation, convergence testing, and HPC scheduling. El Agente Q \cite{el-agente-q} similarly demonstrated a multi-agent cognitive architecture for molecular quantum chemistry, achieving high success rates across geometry optimization, frequency analysis, and spectroscopic property tasks. Liu et al.~\cite{liu_vaspilot_2025} introduce VASPilot, a multi-agent system that automates general VASP workflows including band-structure calculations, convergence tests, and lattice optimizations. While VASPilot demonstrates that agentic orchestration can reliably handle routine, well-defined DFT tasks, it is not designed for the adaptive, closed-loop recovery that TS search demands, and it provides no mechanism for multimodal path diagnosis or mid-workflow replanning when a TS calculation fails. Xia et al. \cite{amc} also demonstrated agentic execution of DFT calculations like structural relaxation, band structure, adsorption energy, and crucially TS search. However, Xia et al. use a predefined workflow library with LLM-driven parameter selection rather than a system capable of dynamic replanning and adaptive recovery outside predefined workflows, leading to a TS search success rate of only about 40\%. In summary, existing agentic DFT frameworks are not equipped to provide the multimodal, physics-grounded diagnostic loop that robust TS search demands. 


Previous work has also explored machine learning approaches for TS search, rather than relying solely on DFT calculations. CatTSunami \cite{ocp-neb} demonstrated that foundation machine-learned interatomic potentials (MLIPs) \cite{equiformer-v2, gemnet, painn, dimenet++}, pretrained on the Open Catalyst Project \cite{oc-20}, can enable zero-shot TS searches. Jung et al. \cite{nebscape} introduce a fully scripted pipeline that automates the construction and ranking of candidate reaction paths before launching MLIP-driven TS searches. Meissner et al. \cite{evolving-ts} use an agentic meta-optimizer to orchestrate MLIP-driven TS search workflows using OpenEvolve. These approaches operate exclusively using MLIPs, where smooth energy surfaces and millisecond-scale force evaluations make the dominant failure modes addressable by static parameter sweeps or pre-NEB heuristics. We target the qualitatively harder DFT-level regime, where each TS search is a multi-hour HPC calculation with diverse failure modes requiring closed-loop, multimodal diagnosis that no static script or pre-evolved program can provide. DFT remains the dominant quantum-mechanical framework for mechanistic studies in catalysis and serves as the primary source of training and validation data for MLIPs.

\section{TSAgent: An Agentic Workflow for Autonomous Transition State Search}
\label{sec:methodology}

\subsection{TSAgent Workflow}

\begin{figure}[t]
    \centering
    \includegraphics[width=0.9\linewidth]{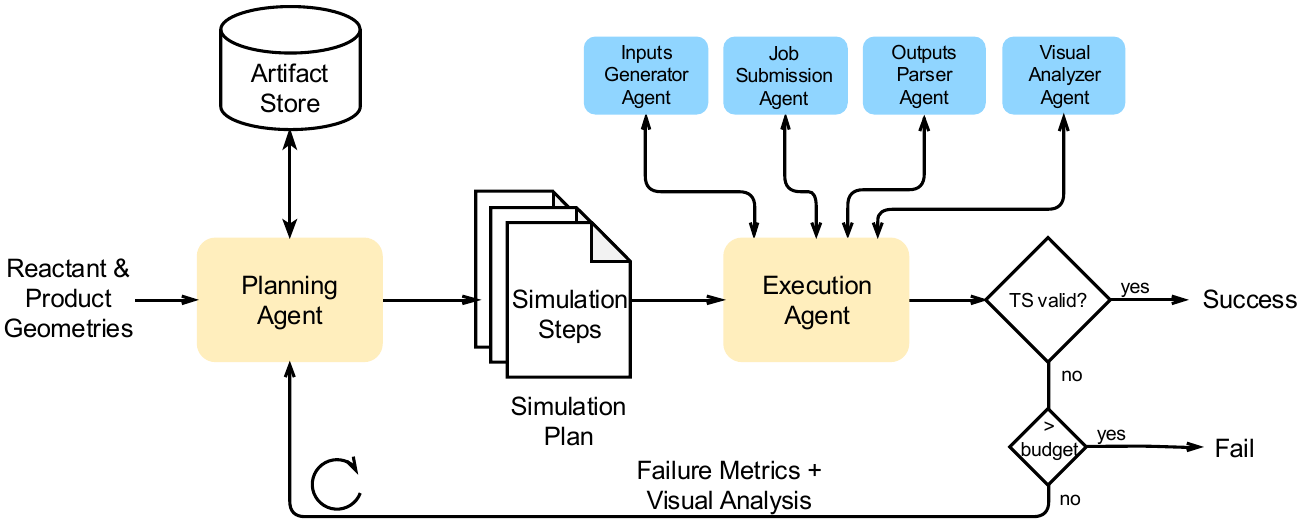}
    \caption{Overview of the TSAgent workflow. Because TS searches are failure-prone and recovery requires physics-informed corrections that vary by failure mode, we structure the workflow as a plan-execute-analyze-replan loop. Given reactant and product geometries, the Planning Agent generates a system-specific \texttt{SimulationPlan}, which is orchestrated by the Execution Agent by interacting with the simulation environment through specialized sub-agents. If any \texttt{SimulationStep} fails, the Planning Agent uses the resulting diagnostics and visual evidence to identify the underlying cause and revise the plan to fix the specific failure mode. The loop continues until a physically validated TS is found, or the compute budget is exhausted.}
    \label{fig:agentic-workflow}
\end{figure}

Given reactant and product geometries (i.e., atomic coordinates and atom types), the goal is to identify the first-order saddle point that connects them along the reaction path. From a systems perspective, finding the TS is not a single simulation task, but rather a long-horizon decision problem with delayed feedback, requiring a complex sequence of long-running DFT calculations on HPC infrastructure. Feedback is delayed and asynchronous, often arriving only after hours or days of wall-clock time. Moreover, intermediate outputs must be interpreted scientifically before the next action can be chosen. To effectively address these requirements, we implement TSAgent as a persistent plan$\rightarrow$execute$\rightarrow$analyze$\rightarrow$replan workflow as shown in Figure \ref{fig:agentic-workflow}. Planning and execution are explicitly decoupled to separate high-level scientific reasoning from low-level interaction with the simulation environment, allowing the Planning Agent (PA) to focus on simulation strategy and the interpretation of evidence while the Execution Agent (EA) handles the operational complexity of interacting with the simulation environment via tools. All agents in the workflow, including the PA, the EA, and the specialized sub-agents introduced below, use GPT-5.4 and are differentiated by their role-specific system prompts and the set of tools they can invoke. Subsequent sections explain each module in detail, with exact prompts and tools in Appendix \ref{app:prompts}.

\subsubsection{Planning Agent (PA)}

To organize the sequence of DFT calculations that need to be executed to find the TS, the PA uses the \texttt{SimulationStep} and \texttt{SimulationPlan} schema. All instructions, settings, and parameters involved in a particular DFT calculation are represented via a schema called a \texttt{SimulationStep}, and the full sequence of \texttt{SimulationSteps} and dependencies required in the current phase are organized into a \texttt{SimulationPlan}. These schema allow the agents to maintain a coherent strategy across multistage searches, step failures, and repeated restarts, while being easily extendable to other DFT calculations. The agent uses three main \texttt{SimulationStep} types:

\textbf{Geometry Optimization (GO):} GO calculations are used to relax a given atomic structure to a nearby local minimum on the potential energy surface. Double-ended TS search algorithms like the Nudged Elastic Band (NEB) algorithm assume that the two endpoints define minima on the same potential energy surface, so GO must be applied to the reactant and product geometries beforehand. 

\textbf{Nudged Elastic Band (NEB):} The TS search algorithm we use herein is the climbing image NEB algorithm \cite{henkelman-ci-neb}. The NEB algorithm constructs a discrete reaction path between reactant and product geometries by introducing a series of intermediate molecular configurations (``images''), each representing the same atoms arranged at different stages along the reaction. Adjacent images are connected by artificial springs to form a band of images. The images are iteratively optimized using forces decomposed into components perpendicular and parallel to the path: the true potential energy gradient drives relaxation toward the minimum-energy path, while spring forces maintain spacing along it. At convergence, the band approximates the minimum-energy path, and the highest-energy image is the candidate TS.

\textbf{Vibrational Frequency Analysis (VFA):} Once a candidate TS image is identified, VFA is performed to verify that it is located at a first-order saddle point. This is achieved by evaluating the Hessian of the potential energy surface at the candidate image, whose eigenvalues determine vibrational frequencies. A minimum has only real frequencies, whereas a TS has exactly one imaginary frequency, and multiple imaginary frequencies indicate higher-order saddles.

Together, these three step types compose typical TS-search workflows, but a \texttt{SimulationPlan} is rarely a single linear GO--NEB--VFA sequence. A difficult search may, for instance, require phased convergence with looser-then-tighter NEB restarts, repeated geometry optimizations of the endpoints, or a return to an earlier NEB stage when VFA yields zero or multiple imaginary frequencies. In more complex cases the planner must even revise the problem decomposition itself, splitting a pathway that contains a stable intermediate image minimum into independent child TS searches over the resulting sub-reactions. The \texttt{SimulationStep} and \texttt{SimulationPlan} schema therefore serve as the interface between domain-level chemical reasoning and executable DFT workflows, exposing each algorithm to the PA as an interchangeable scientific operator rather than a hard-coded script. Although the present implementation focuses on NEB-based TS search, the abstraction is intentionally algorithm-agnostic: alternative methods such as the dimer method \cite{henkelman_dimer_1999}, eigenvector-following \cite{sharada_finite_2014, hermes_accelerated_2019}, or growing-string algorithms \cite{string-mina2017} can be incorporated by adding new \texttt{SimulationStep} types with their own typed inputs, outputs, and failure signatures. Building on these abstractions, the PA receives the reactant and product geometries along with any user-provided context, reasons over the chemistry of the system, and returns a tailored \texttt{SimulationPlan} that the EA then operationalizes.

\subsubsection{Execution Agent (EA)}

The EA executes the generated \texttt{SimulationPlan} step-by-step, interfacing with the simulation environment through sub-agents to run DFT calculations using the Vienna Ab initio Simulation Package (VASP) \cite{kresse_efficient_1996}. Using the simulation parameters identified in the \texttt{SimulationStep}, the input-generation sub-agent prepares the VASP simulation directory with the required inputs and configuration files. The job-submission sub-agent interacts with the specific HPC environment to submit and monitor jobs. Since DFT calculations may take multiple hours to complete, the workflow persists its state and pauses execution while the jobs are running, and restarts once the jobs have completed (either completing successfully or terminating with an error). 

The outputs parser extracts structured diagnostics from VASP output files using a collection of pattern-based extractor tools, including final energies and per-image energy profiles, maximum and root-mean-squared ionic forces, convergence flags, imaginary vibrational frequency counts and frequency values, and warning signatures. We use pattern matching here rather than having the agent directly parse the outputs, because a single DFT calculation produces several output files spanning thousands of lines of low-level numerical data, and routing this raw output directly to an agent would both overflow its context window and introduce a significant hallucination risk over long multi-step analysis. While these scalar diagnostics are sufficient to detect most simple failure modes, certain pathway pathologies only become apparent through inspection of the 3D atomic configurations themselves, such as atom collisions or sudden non-physical atomic rearrangements or motions between adjacent NEB images. To recover this missing channel of information, a dedicated visual analyzer agent renders orthogonal projections of the reaction pathway and examines them in much the same way a human DFT practitioner would when debugging a suspect NEB run, returning structured qualitative observations that augment the output parser's numerical report. 

The extracted diagnostics are used by the EA to determine next steps. If the diagnostics indicate that the current \texttt{SimulationStep} completed successfully, the EA advances to the subsequent step in the plan. If the diagnostics indicate a failure, the EA summarizes the failure and supporting evidence and defers to the PA for replanning. We explain how the PA fuses visual and numerical signals to generate a revised simulation strategy in the following section.

\subsubsection{Multimodal Diagnosis and Adaptive Replanning}

\begin{figure*}[t]
    \centering
    \includegraphics[width=\linewidth]{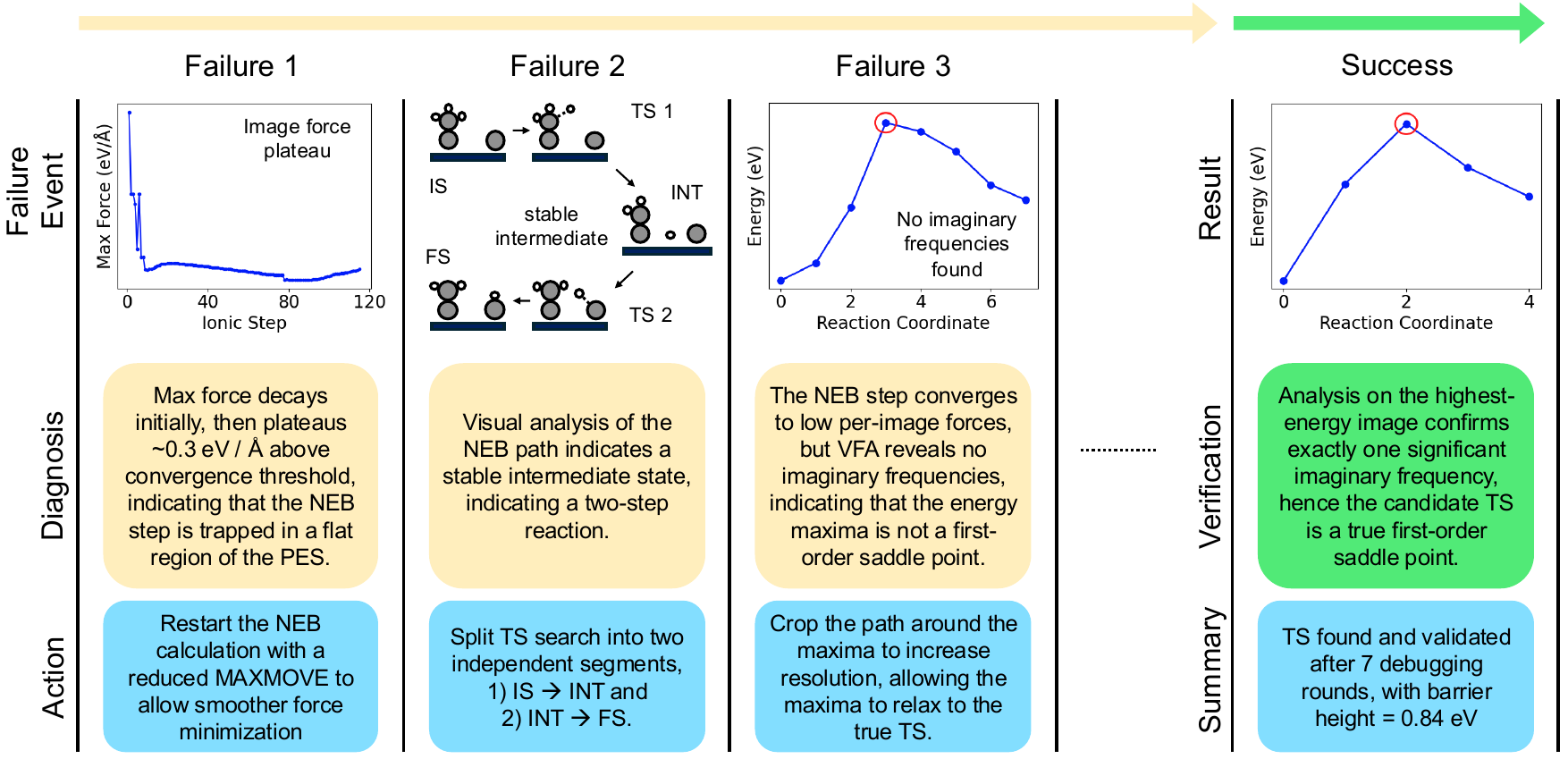}
    \caption{Illustration of iterative diagnosis and adaptive replanning in TSAgent. Each iteration combines intermediate results, or failure signals, with numerical and visual evidence to produce a structured diagnosis and a corresponding corrective action. The final stage distinguishes workflow completion from scientific verification by requiring physics-based validation of a true first-order saddle point before declaring success.}
    \label{fig:sim-log}
\end{figure*}

If a failure is identified during execution of the current plan, the PA diagnoses the issue using the failure summary provided by the EA, diagnostics from prior steps, expert-crafted debugging guidelines, and a replan log recording previous replanning decisions and their outcomes. Using this information, the PA either generates a revised plan to fix the failure, or escalates to the user if the error cannot be resolved. Figure~\ref{fig:sim-log} illustrates a complete diagnosis and replanning episode across successive rounds of a hypothetical reaction pathway, each exposing a qualitatively distinct failure mode. We use this example to explain the replanning process. 

In straightforward cases, the structured diagnostics extracted by the output parser are sufficient to identify the failure mode and prescribe an intervention. Panel~1 of Figure~\ref{fig:sim-log} illustrates this: the NEB maximum force curve decays rapidly before plateauing at approximately $0.3$~eV / \AA{} above the convergence threshold and remains stuck for more than 120 ionic steps, indicating that the optimizer is trapped in a flat region of the PES. This means that the atomic displacements per step are large enough that the optimizer continually overshoots the shallow basin, analogous to a gradient descent run with an excessively large step size in a flat loss landscape. To fix this issue, the agent proposes restarting from the current NEB geometries with a reduced \texttt{MAXMOVE}---the parameter governing the maximum displacement any atom may undergo in a single ionic step---directly from the numeric signal, and no visual inspection of the pathway is required.

Panel~2 illustrates a case where numeric diagnostics are insufficient. In this step, a visual analysis of the NEB pathway indicated the H atom first dissociated from the adsorbed $^{*}\mathrm{C_2H_3}$ complex, then transiently formed a stable surface-adsorbed intermediate ($^{*}\mathrm{H}$), before bonding with an adsorbed $^{*}\mathrm{C}$ atom. This indicates that the process is not a simple transfer reaction, but rather a multi-step reaction proceeding through two elementary steps with two separate TSs. In this case, the agent splits the TS search into two independent TS searches---from the reactant to the stable intermediate, and from the intermediate to the final product state. This kind of latent problem decomposition, triggered by evidence gathered mid-execution, highlights the need for an agentic workflow over a static workflow or parameter-sweep script, and closes a diagnostic gap that has historically required a human practitioner to inspect rendered molecular geometries. Together, numeric and visual diagnostics form a complementary feedback loop: the workflow iterates through successive diagnosis-and-replan cycles, drawing on whichever modality is informative for the current failure mode, until either a theoretically validated TS is recovered or a user-specified compute budget is exhausted and the TS search is escalated. If the workflow reports that a TS has been found, the agent performs a final validation by checking whether the TS search algorithm converged to the user-specified thresholds, whether VFA identifies exactly one significant imaginary frequency along the reaction coordinate, and if the barrier height was energetically consistent with the reactant and product geometries. Only when all three criteria are jointly satisfied is the TS declared validated, as illustrated in Panel~4.

\section{Experiments}

To study the accuracy and scientific applicability of TSAgent, we evaluate its performance in three settings. First, we benchmark TSAgent on a subset of the OC20NEB dataset, showing that it can reliably recover TSs across catalyst surfaces, elements, adsorbates, and reaction classes. Second, we compare TSAgent against expert DFT practitioners on a held-out set of OC20NEB reactions, finding that TSAgent can match human-level performance while reducing manual intervention. Finally, we deploy TSAgent in a realistic scientific setting by using it to reproduce Br{\o}nsted--Evans--Polanyi scaling relationships for NH$_3$ dissociation on various catalyst surfaces.



\subsection{Evaluating TSAgent on the OC20NEB Heterogeneous Catalysis Transition States Dataset}
\label{sec:OC20NEB_eval}
\label{sec:OC20NEB_results}

\textbf{Experiment Setup.} We evaluate TSAgent on the OC20NEB heterogeneous catalysis transition states dataset ~\cite{ocp-neb}. OC20NEB contains three reaction types---dissociation, transfer, and desorption. However, due to compute constraints (each TS search requires about 10000 cpu-hours on average with our settings), we limit our analysis to a stratified 100 reaction subset of dissociation and transfer reactions. We exclude desorption reactions from our analysis because the majority of them are barrierless and hence have no TS ~\cite{ocp-neb}. The selected subset nonetheless spans 100 distinct catalyst surfaces, 51 elements, 29 adsorbate types, and 8 bond-breaking reaction types. We allow a budget of five debugging (replanning) rounds per TS search. It should be noted that DFT total energies are sensitive to the precise choice of exchange-correlation functional, k-point sampling, and convergence parameters, such that even small residual differences in setup can shift absolute TS energies regardless of whether the correct saddle point was found. Because we cannot guarantee that our settings are identical to those used to construct OC20NEB across all such parameters, we refrain from direct numeric comparisons of TS energies against the dataset ground truth, and instead evaluate success through physics-based structural criteria. A full description of the reactions chosen, the DFT settings applied, and TSAgent instructions are found in Appendix \ref{app:oc20neb-cases}. For this dataset, we compare TSAgent against an MLIP-based TS search pipeline using the Universal Model for Atoms (UMA) ~\cite{uma} (\texttt{uma-s-1p2}) as the energy and force calculator. The pipeline follows the two-stage climbing-image NEB protocol of CatTSunami ~\cite{ocp-neb}, replacing the older Equiformer-V2 model \cite{equiformer-v2} with the more recent UMA foundation MLIP. Full implementation details are provided in Appendix~\ref{app:uma_baseline}.


\begin{figure}[t]
  \centering 
  \includegraphics[width=1\linewidth]{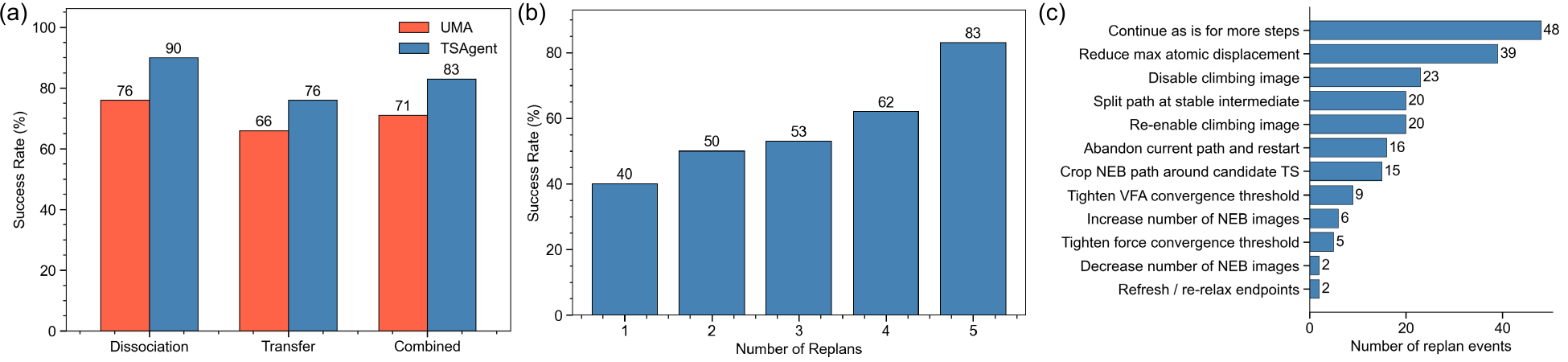}
    \caption{TSAgent performance on a diverse subset of the OC20NEB heterogeneous catalysis transition states benchmark. \emph{(a)} TSAgent improves over the UMA baseline across both reaction classes. \emph{(b)} TSAgent success rate scales steadily with the replan budget, showing that iterative adaptation rather than any one-shot fix is what drives the improvement. \emph{(c)} The planner exercises a wide range of replan actions, indicating the gains arise from genuine policy diversity rather than static interventions.}
  \label{fig:oc20neb_results}
\end{figure}

\textbf{Results.} TSAgent achieves an overall success rate of 83\% on the benchmark (95\% CI via Wilson score: 74.5--89.1\%; Figure~\ref{fig:oc20neb_results}a), outperforming the UMA baseline. Dissociation TSs were found in 90\% of the cases (95\% CI: 78.6--95.7\%), whereas hydrogen transfer (H-transfer) tasks proved more challenging with a success rate of 76\% (95\% CI: 62.6--85.7\%). This gap is consistent with the mechanistic complexity of H-transfer reactions, which often require the identification of a concerted bond-breaking and bond-forming pathway in which both the donor and acceptor geometries must be simultaneously satisfied at the TS. Furthermore, Figure~\ref{fig:oc20neb_results}b shows that the one-shot success rate, achieved without any replanning, stands at only 40\%, with gains accruing incrementally across subsequent replan rounds. This trajectory underscores that robust TS discovery depends on iterative, evidence-driven adaptation rather than any single well-chosen initial strategy. Figure~\ref{fig:oc20neb_results}c highlights the specific interventions applied by the planner across all replan events for successful runs. Each replan event can combine more than one action (e.g., first disabling the climbing image algorithm, then reducing the maximum atomic displacement at a later step). In addition to the policy diversity, several of the interventions involve tuning exact parameter values, adding an additional layer of complexity. The fact that success draws on this full repertoire rather than a single simplistic tweak supports our hypothesis that the planner is matching interventions to evidence on a per-case basis, consistent with the iterative-adaptation picture in Figure~\ref{fig:oc20neb_results}b.

\subsection{Benchmarking TSAgent Against Human Experts}
\label{sec:human-expert-benchmark}

\textbf{Experimental Setup.} We compare TSAgent against three human experts on 10 OC20NEB reactions (5 dissociation and 5 transfer reactions, distinct from the reactions in Section~\ref{sec:OC20NEB_eval}). The experts are PhD students that have between 3 to 5 years of experience running TS calculations with DFT, and each have multiple publications in peer-reviewed computational catalysis journals. We compare their performance on three metrics: success rate (SR), cpu-hours, and operator effort. Operator effort captures the manual burden of inspecting structures, diagnosing failed optimizations, modifying input files, resubmitting jobs, and deciding whether a saddle-point candidate warrants vibrational frequency analysis. For TSAgent, the corresponding figure is the initialization, monitoring, and final verification time. Both the experts and the agent were instructed to use identical DFT settings and pre-defined success criteria. A full description of the chemistry of the chosen reactions, DFT settings, and standard operating procedures can be found in Appendix \ref{si:sec:human-expert-benchmark}.

\textbf{Results.} Table~\ref{tab:human_aggregate_results} summarizes the aggregate performance for TSAgent, each human expert (HE01--HE03), and the average for human expert. TSAgent solved 70\% cases overall, compared to a human-expert average of $73 \pm 12$\% and a best-human result of 80\%. Human experts spent on average 47~min per successful case on debugging and correction. At the scale of a real mechanistic study---where a practitioner might execute 50 to 100 TS searches across a reaction network---this per-case burden compounds into weeks of focused manual effort, creating a practical ceiling on the scope of investigations that a single researcher or small group can conduct. TSAgent reduces this monitoring overhead regardless of search complexity, making large-scale mechanistic campaigns tractable without additional personnel. However, TSAgent consumed approximately 3100 more cpu-hours per case than the human-expert average on average. While a human expert can sometimes identify the dominant failure mode from a single visual inspection and act on it directly, TSAgent requires additional iterations before converging on a winning strategy. A detailed comparison of the performance on each case, including the predicted TS energies, is shown in Appendix \ref{si:sec:human-expert-benchmark}.

\begin{table}[t]
\centering
\normalsize
\setlength{\tabcolsep}{4pt}
\renewcommand{\arraystretch}{1.12}
\caption{Comparison of TSAgent against three human experts (HE01--HE03) on 10 OC20NEB reactions across three metrics; success rate (SR), average cpu-hours, and average operator efforts. CPU-hours and operator time are computed over successful cases only. TSAgent matches human-expert success rates without significant operator effort, demonstrating that agentic TS search can substitute for manual expert intervention.}
\label{tab:human_aggregate_results}
\begin{tabularx}{\linewidth}{%
    >{\raggedright\arraybackslash}p{0.33\linewidth}
    >{\centering\arraybackslash}X
    >{\centering\arraybackslash}X
    >{\centering\arraybackslash}X
    >{\centering\arraybackslash}p{0.17\linewidth}
    >{\centering\arraybackslash}X}
\toprule
\textbf{Metric}
    & \textbf{HE01} & \textbf{HE02} & \textbf{HE03}
    & \textbf{HE-average} & \textbf{TSAgent} \\
\midrule
Dissociation SR (\%)                 & 80   & 80   & 80   & $80 \pm 0$        & 40 \\
Transfer SR (\%)                     & 80   & 80   & 40   & $67 \pm 23$       & 100 \\
Overall SR (\%)                      & 80   & 80   & 60   & $73 \pm 12$       & 70 \\
Average cpu-hours                    & 8556 & 4825 & 7418 & $6708 \pm 1912$   & 9808 \\
Average operator efforts (min)       & 63   & 51   & 31   & $47 \pm 16$       & -  \\
\bottomrule
\end{tabularx}
\end{table}


\subsection{Reproducing Br{\o}nsted--Evans--Polanyi (BEP) Relationship for NH$_3$ Dissociation on Pure Metals and SAAs}
\label{sec:published-validation}
 
Beyond benchmark reactions drawn from curated datasets, we evaluate TSAgent's ability to reproduce Br{\o}nsted--Evans--Polanyi (BEP) scaling relationships for ammonia (NH$_3$) dissociation (NH$_3^*$ $\rightarrow$ NH$_2^*$ + H$^*$) on transition metals and alloys \cite{darby_elucidating_2018}. These relations, which remain one of the most fundamental and impactful concepts in heterogeneous catalysis \cite{wiley_fundamental_2014}, show that the reaction energy (i.e., $\Delta H_{\text{rxn}} = E_{\text{NH}_2^* + \text{H}^*} - E_{\text{NH}_3^*}$) scales linearly with the activation barrier ($E_{\text{A}} = E_{\text{NH}_3^* \cdots \text{H}} - E_{\text{NH}_3^*}$). Therefore, the reaction energy can be used as an approximation for the activation barrier of the reaction, especially for new, unstudied catalysts. Darby \emph{et al.} \cite{darby_elucidating_2018} demonstrated that single atom alloys (SAAs) (i.e., when one metal is doped onto the surface of another in dilute concentrations) follow different scaling relations than pure metals, indicating that a single universal scaling relation does not apply across both material classes. The validation objective is therefore to determine whether TSAgent can recover, across all surfaces, a consistent set of TSs that reproduce both the variation in activation barriers and the separation between the pure-metal and SAA BEP trends. 

\begin{figure}[t]
  \centering
  \includegraphics[width=0.85\linewidth]{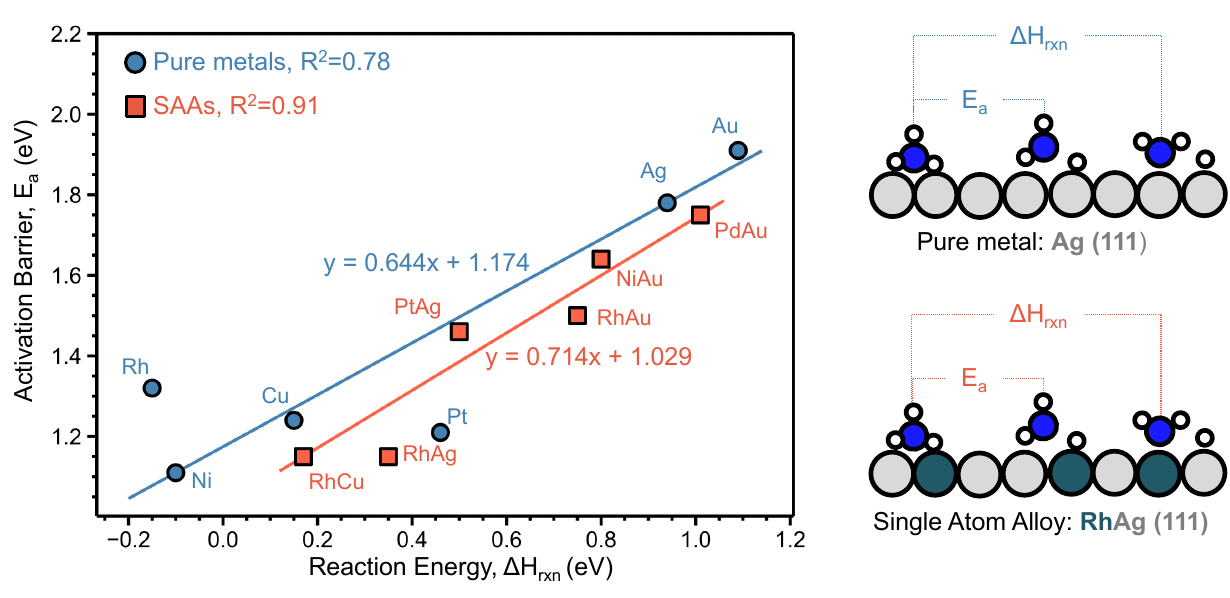}
  \caption{Br{\o}nsted--Evans--Polanyi (BEP) relationships as computed by TSAgent for NH$_3$ dissociation
  (NH$_3^* \rightarrow$ NH$_2^*$ + H$^*$) on pure metal (blue) and SAA (red) surfaces, where the activation energy $E_\mathrm{a}$ is plotted against reaction energy $\Delta E_\mathrm{rxn}$. The linear BEP fits are shown for each class.}
  \label{fig:NH3-casestudy}
\end{figure}

We use TSAgent to compute the activation barrier and reaction energy for 12 pure metal and SAA catalysts, and compile the results into BEP scaling relations shown in Figure \ref{fig:NH3-casestudy}. Consistent with Darby \emph{et al.}~\cite{darby_elucidating_2018}, the activation barrier correlates strongly with the reaction energy with R$^2$ = 0.78 for pure metals (blue) and R$^2$ = 0.91 for SAAs (red). However, Figure~\ref{fig:NH3-casestudy} reiterates that the SAA scaling relation is indeed different from the pure metals, where the scaling relations differ primarily in their intercept (1.174 for pure metals vs 1.029 for SAAs). Both Pt and Rh catalysts deviate significantly from their scaling relations. While the deviation in Pt was reported in the original work, we attribute the Rh deviation, along with other differences between the scaling relations to the less computationally expensive exchange-correlation functional we use versus Darby \emph{et al.}~\cite{darby_elucidating_2018}. 

\section{Conclusion}
\label{sec:conclusion}

In this work, we presented TSAgent, an agentic workflow for autonomous transition-state (TS) search using first-principles quantum chemistry calculations based on density functional theory (DFT). By decoupling high-level scientific reasoning from operational interaction with the simulation environment, and by closing the loop with both numerical and visual diagnostics, TSAgent handles the long-horizon, asynchronous, multimodal feedback that characterizes quantum mechanics based TS searches that earlier agentic frameworks for atomistic simulations were not designed to address. To demonstrate the effectiveness and scientific applicability of TSAgent, we evaluated it on a subset of the OC20NEB heterogeneous catalysis transition states dataset, compared its performance against human experts, and used it to reproduce Brønsted–Evans–Polanyi scaling relationships from a published heterogeneous catalysis study. 

By automating the TS search step, TSAgent has the potential to accelerate mechanistic studies and the generation of reaction-level training data. In doing so, it can accelerate populating the off-equilibrium regions of chemical space that current datasets leave underrepresented~\cite{ocp-neb, transition1x}, providing saddle-region supervision needed by the next generation of reactive interatomic potentials and generative TS models~\cite{diff-ts-generation, ot-ts-generation}. While this work demonstrates the TS search workflow for heterogeneous catalysis, the underlying approach is broadly applicable across chemical domains.  

\textbf{Limitations.}
\label{sec:limitations}
We highlight some key limitations of TSAgent. First, the workflow assumes that reasonable reactant and product geometries are provided as input. However, generating these geometries is itself a nontrivial problem that must be solved for large-scale mechanistic studies. Second, due to the substantial compute cost of DFT-level TS searches, our OC20NEB evaluation is restricted to 100 reactions sampled from the total 600 reactions available. Finally, the dominant cost of TSAgent remains the DFT calculations themselves rather than LLM inference: a successful search consumes thousands of cpu-hours, and each replan extends this further. This is consistent with the cost expert practitioners pay today, but it limits throughput in high-volume mechanistic screening campaigns. Additionally, while TSAgent performs a rigorous multi-criteria validation before declaring success, LLM-based verification is not a physical guarantee, and automated TS assignments accepted without any human oversight could propagate errors into downstream mechanistic conclusions or into reaction-level training datasets.


\clearpage

\bibliographystyle{unsrtnat}
\bibliography{references}

\clearpage

\appendix

\section*{Appendix}

\section{Software Stack and LLM Backbone}
\label{app:software-stack}

All agents in the system use OpenAI's \texttt{gpt-5.4} as the backbone language model, accessed through the OpenAI Agents SDK. We use a uniform backbone across all roles rather than mixing model families; differentiation between agents is achieved through the reasoning effort budget rather than model size. The Planning Agent and Execution Agent run with \texttt{reasoning effort = high}, the Visual Analyzer Agent with \texttt{effort = medium}, and the three stateless sub-agents (Inputs Generator, Job Submission, Outputs Parser) with \texttt{effort = low}. This tiering keeps cost proportional to the deliberation each role actually requires: the planner and executor must reason over multi-step trajectories and failure evidence, while the sub-agents execute narrow, well-defined procedures.

Agent orchestration is implemented with a custom \texttt{AgentWrapper} class built on top of the OpenAI Agents SDK. We chose a direct SDK integration rather than a higher-level framework such as LangGraph because the control flow in our system is explicitly managed by the orchestrator: the planner generates a typed \texttt{SimulationPlan}, the executor iterates over it deterministically, and the replan trigger is a hard schema boundary rather than an emergent routing decision. A graph-based abstraction would add indirection without simplifying any of these transitions. Agent conversation state is persisted in a SQLite database with one session per planning or execution invocation.

DFT calculations are performed with VASP 6.3. Structure preparation, NEB path interpolation, and file I/O use ASE 3.26.0. 

\section{HPC Environment and Asynchronous Execution}
\label{app:hpc}

All DFT calculations are submitted to a SLURM-managed institutional HPC cluster using exclusive 128-core nodes (partition \texttt{RM}). Each job is submitted via a pre-configured shell script that calls \texttt{sbatch} with a default wall-clock limit of 16 hours per job. The number of MPI tasks per node is computed at runtime by the Execution Agent using a \texttt{calculate\_vasp\_parallel\_params} tool that selects \texttt{NCORE} based on the number of NEB images and available cores.

The system is designed around the constraint that VASP jobs can run for hours to days, far beyond any LLM context window. Execution is therefore checkpointed after every job submission. All persistent state is written as JSON or JSONL files under a per-run \texttt{artifacts/} directory:

\begin{itemize}[leftmargin=0.5cm]
  \item \texttt{runtime\_state.json} --- current run status and plan index, updated atomically after every state transition.
  \item \texttt{current\_plan\_status.json} --- per-step status records, including parser diagnostics and artifact pointers.
  \item \texttt{plan\_store.jsonl} --- append-only log of every \texttt{SimulationPlan} generated across all planning invocations.
  \item \texttt{replan\_log.jsonl} --- append-only log of \texttt{ReplanDecision} objects, one per replanning cycle.
  \item \texttt{failure\_events.jsonl} --- structured \texttt{FailureEvent} objects emitted by the Execution Agent.
  \item \texttt{.job\_monitor/jobs/<job\_id>.json} --- per-SLURM-job records written by the job monitor.
\end{itemize}

A cron-based job monitor polls SLURM via \texttt{squeue} at a configurable interval. When all tracked jobs for the current plan step have left active states (\texttt{RUNNING}, \texttt{PENDING}, \texttt{COMPLETING}), the monitor invokes \texttt{main.py restart}, which reloads \texttt{runtime\_state.json} and resumes the orchestrator from the exact point at which it last checkpointed. This design means the orchestrator process is entirely stateless between runs: every restart reconstructs full context from the artifact store, and no in-memory state survives across SLURM job boundaries.

\section{Agent Architecture and Schemas}
\label{app:schemas}

\subsection{SimulationStep Schema}

The \texttt{SimulationStep} is the fundamental typed unit of work passed from the Planning Agent to the Execution Agent. All three step types (GO, NEB, VFA) inherit from a common base that captures the execution directory, KPOINTS grid, and INCAR parameters; type-specific fields encode the additional context each calculation requires.

\begin{lstlisting}[language=Python, basicstyle=\small\ttfamily,
    commentstyle=\color{gray}\itshape, columns=fullflexible]
# --- Shared base fields (all step types) ---
step_name: str           # Unique identifier within the plan
step_type: Literal[      # Discriminator for runtime dispatch
    "GeometryOptimization",
    "NudgedElasticBand",
    "VibrationalFrequencyAnalysis"
]
step_INCAR: BaseSimulationINCAR   # VASP INCAR parameters (typed per step)
step_KPOINTS: List[int]  # Gamma-centered k-point grid, e.g. [2, 2, 1]
step_execution_directory: str     # Absolute path where VASP inputs/outputs live

# --- GeometryOptimizationStep extra fields ---
initial_geometry_path: str        # Input POSCAR / CONTCAR for this relaxation
restart_from: Optional[str]       # Previous GO directory to warm-start from

# --- NudgedElasticBandStep extra fields ---
reactant_geometry_path: str       # Relaxed IS geometry (CONTCAR from GO, or raw POSCAR)
product_geometry_path: str        # Relaxed FS geometry
num_images: int                   # Number of internal NEB images (must match INCAR IMAGES)
move_threshold: float             # Angstrom threshold for spectator-atom alignment
restart_from: Optional[str]       # Previous NEB directory to warm-start from

# --- VibrationalFrequencyAnalysisStep extra fields ---
imaginary_frequency_threshold: float  # Minimum |frequency| (meV) to treat as imaginary
candidate_TS_image_path: Optional[str]  # Set by EA from NEB output; planner leaves None
\end{lstlisting}

The \texttt{INCAR} objects are themselves typed Pydantic models (\texttt{GO\_INCAR}, \texttt{NEB\_INCAR}, \texttt{VFA\_INCAR}) that enforce step-specific constraints, e.g.\ \texttt{IBRION = 2} for geometry optimization, \texttt{IBRION = 3} for NEB, and \texttt{IBRION = 5} for vibrational analysis. Pydantic validators additionally enforce consistency rules such as \texttt{NEB\_INCAR.IMAGES == NudgedElasticBandStep.num\_images} and the requirement that \texttt{restart\_from}, when provided, must be a different directory from \texttt{step\_execution\_directory}.

\subsection{SimulationPlan Schema}

\begin{lstlisting}[language=Python, basicstyle=\small\ttfamily,
    commentstyle=\color{gray}\itshape, columns=fullflexible]
class SimulationPlan:
    plan_id: str                    # e.g. "PLAN_1", "PLAN_2"; incremented per replanning cycle
    simulation_steps: List[SimulationStep]   # Ordered sequence; EA executes left to right
\end{lstlisting}

Step ordering encodes the dependency graph implicitly: later steps reference file paths written by earlier steps (e.g.,\ a NEB step's \texttt{reactant\_geometry\_path} points to the \texttt{CONTCAR} produced by a prior GO step). There is no explicit dependency field; the Planning Agent is responsible for ensuring that path references are consistent across steps, and the Execution Agent validates that referenced paths exist before each input-generation call.

Warm restarts are encoded through \texttt{restart\_from}: rather than continuing in the same directory, the plan allocates a new \texttt{step\_execution\_directory} and provides the completed step's directory as \texttt{restart\_from}. The Inputs Generator Agent then copies checkpoint files (\texttt{CONTCAR}, \texttt{WAVECAR} when available) into the new directory before regenerating \texttt{INCAR}, \texttt{KPOINTS}, and \texttt{POTCAR}. This keeps every execution attempt in its own named directory, which simplifies auditing and avoids overwriting prior outputs.

Multi-step decompositions (child TS searches spawned when an intermediate stable state is detected) are launched as independent runs via a \texttt{spawn\_multi\_step\_ts\_search} tool. Each child run receives its own \texttt{run\_id} and operates as a fully independent orchestrator instance; the parent run records child run IDs in its artifacts but does not manage their execution.

\subsection{FailureEvent and StepStatus Schemas}

\begin{lstlisting}[language=Python, basicstyle=\small\ttfamily,
    commentstyle=\color{gray}\itshape, columns=fullflexible]
# FailureMetadata: evidence attached to a FailureEvent
class FailureMetadata:
    key_metrics: Dict[str, List[float]]  # Named numeric time-series (forces, energies, etc.)
    visual_summary: Optional[NEBVisualSummary]  # Structured output from visual debugger, if run
    artifact_paths: Dict[str, str]       # Pointers to rendered plots and structure previews
    observations: Optional[str]          # Human-readable summary of the failure evidence

# FailureEvent: structured failure report from EA to PA
class FailureEvent:
    step_name: str                   # Which step in the plan failed
    step_execution_directory: str    # Where VASP outputs can be inspected
    failure_step_type: Literal[      # Coarse step category
        "PreflightTests", "GeometryOptimization",
        "NudgedElasticBand", "VibrationalFrequencyAnalysis"
    ]
    failure_step_stage: Literal[     # Where in the step lifecycle failure occurred
        "PREFLIGHT", "INPUT_GENERATION", "JOB_SUBMISSION",
        "RUNTIME", "OUTPUT_ANALYSIS", "VALIDATION"
    ]
    failure_reason: str              # Free-text description of the specific failure
    evidence: List[FailureMetadata]  # One entry per failed step; carries numeric diagnostics
\end{lstlisting}

\begin{lstlisting}[language=Python, basicstyle=\small\ttfamily,
    commentstyle=\color{gray}\itshape, columns=fullflexible]
# StepStatus: per-step execution record maintained by the EA
class StepStatus:
    step_id: str                     # Matches step_name in the plan
    step_type: SimulationStepType    # GO / NEB / VFA
    status: Literal[                 # Lifecycle state of this step
        "TBD", "PAUSED_WAITING_FOR_JOBS", "COMPLETED", "FAILED"
    ]
    step_execution_directories: List[str]  # Directories used (>1 if restarted)
    job_ids: List[str]               # SLURM job IDs submitted for this step
    observations: Optional[str]      # Parser summary: convergence, warnings, metrics
    warning_signature_candidates: List[str]   # Non-fatal signatures (e.g. near-miss forces)
    failure_signature_candidates: List[str]   # Fatal signatures (e.g. NEB_EXPLODING_IMAGES)
    visual_summary: Optional[NEBVisualSummary]  # Visual debugger output, NEB steps only
    artifact_paths: Dict[str, str]   # Rendered artifact paths for this step
    error_message: Optional[str]     # Error detail when status == "FAILED"
\end{lstlisting}

The replan log entry (\texttt{ReplanDecision}) captures what the Planning Agent changed and why:

\begin{lstlisting}[language=Python, basicstyle=\small\ttfamily,
    commentstyle=\color{gray}\itshape, columns=fullflexible]
class ReplanDecision:
    replan_summary: str              # Exact changes made (no explanations)
    replan_rationale: str            # Why the changes are expected to resolve the failure
    restart_mode: Optional[Literal[  # High-level characterization of the intervention
        "CONTINUE_STEP_AS_IS",              # More steps, same parameters
        "RESTART_STEP_WITH_PARAMETER_CHANGE",  # Same starting point, modified INCAR
        "RESTART_FROM_DIFFERENT_STEP"       # Roll back to an earlier step
    ]]
\end{lstlisting}

Each \texttt{ReplanDecision} is appended to \texttt{replan\_log.jsonl} and provided in full to the Planning Agent on every subsequent invocation. This allows the planner to avoid repeating interventions that have already been tried.

\subsection{Outcome Codes}

The Execution Agent returns one of four status codes at the end of each invocation:

\begin{itemize}[leftmargin=0.5cm]
  \item \textbf{PAUSED\_WAITING\_FOR\_JOBS.} The EA has submitted one or more SLURM jobs and is suspending execution. The orchestrator checkpoints current step statuses and exits; the job monitor will restart execution once all tracked jobs leave active SLURM states.

  \item \textbf{FAILED\_AT\_STEP.} A step has failed and the failure is not self-recoverable. The EA populates a \texttt{FailureEvent} via \texttt{build\_failure\_event\_from\_parser\_output} and returns it alongside the output. The orchestrator immediately invokes the Planning Agent in \texttt{REPLAN\_STEP\_FAILED} mode.

  \item \textbf{REPLAN\_REQUESTED.} A step completed without a hard DFT failure, but the output diagnostics indicate that the current plan should not continue as-is. The canonical trigger is \texttt{NEB\_INTERMEDIATE\_MINIMUM}: the NEB converged, but an interior minimum in the energy profile signals a multi-step reaction mechanism. The EA collects visual debugger evidence, constructs a \texttt{FailureEvent}, and returns \texttt{REPLAN\_REQUESTED}. The orchestrator invokes the Planning Agent in \texttt{REPLAN\_REQUESTED} mode.

  \item \textbf{ALL\_STEPS\_SUCCESS.} Every step in the current \texttt{SimulationPlan} has been completed and parsed successfully. The orchestrator passes control to the Planning Agent in \texttt{REPLAN\_ALL\_STEPS\_COMPLETED} mode for final TS validation.
\end{itemize}

\section{Agent Prompts and Tools}
\label{app:prompts}

\subsection{Planning Agent}

The Planning Agent receives a global preamble (describing the simulation environment and the \texttt{SimulationStep} type hierarchy) prepended to its agent-specific instructions at initialization. At runtime, the full \texttt{PlanningAgentInput}---including the current plan, failure event, replan log, per-step diagnostics, and path to the debugging guidelines---is serialized to JSON and passed as the user message.

\begin{promptbox}[Planning Agent System Prompt (Initial Plan and Replanning)]
\small
\textbf{Role.} You are the Planning Agent in this multi-agent system for finding transition states in heterogeneous catalysis reactions. You possess deep domain expertise in DFT, surface science, and statistical mechanics, equivalent to a senior PhD candidate or postdoctoral researcher in Chemical Engineering.

\textbf{Task.} Generate a step-by-step \texttt{SimulationPlan} to find the transition state between the provided IS and FS geometries. Your plan will be executed by the Execution Agent. On failure, generate a revised plan based on the provided failure event, replan history, and debugging guidelines. On successful completion of all steps, validate whether the TS has been found and report the result.

\textbf{Standard workflow.} (1) Geometry optimization of IS and FS (optional if already relaxed). (2) A two-phase NEB: coarse phase with ENCUT 250\,eV and EDIFFG $-$0.1\,eV/\AA\ to establish the path, followed by a refined phase at ENCUT 350\,eV and EDIFFG $-$0.05\,eV/\AA\ seeded from the coarse result. (3) Vibrational frequency analysis on the highest-energy NEB image to confirm a single imaginary mode. The ENCUT and k-mesh used in the final NEB phase must match those used in the GO steps.

\textbf{Replanning.} On each replan invocation you receive the current plan, a structured \texttt{FailureEvent} with numeric evidence and optional visual debugger output, the full replan log, current per-step diagnostics, and debugging guidelines. Refer to the guidelines and combine them with your own domain expertise to generate the revised plan. Do not repeat interventions that appear in the replan log for the same initial/final state pair.

\textbf{Planning modes.}
\texttt{NEW\_PLAN}: generate the first plan.
\texttt{REPLAN\_STEP\_FAILED}: a step reported hard failure.
\texttt{REPLAN\_REQUESTED}: execution requested a replan (e.g.,\ intermediate minimum detected).
\texttt{REPLAN\_ALL\_STEPS\_COMPLETED}: all steps succeeded; validate the TS or escalate.

\textbf{Output.} Return a \texttt{PlanningAgentOutput} with status \texttt{PLAN\_READY}, \texttt{TS\_FOUND}, or \texttt{ESCALATE}.
\end{promptbox}

The Planning Agent has access to the following tools: \texttt{read\_file} and \texttt{list\_files} (read-only access to the run directory for inspecting outputs), \texttt{extract\_elements\_from\_poscar} (determines LMAXMIX from element composition), \texttt{get\_previous\_simulation\_plan} (retrieves earlier plans from the plan store for reference), \texttt{spawn\_multi\_step\_ts\_search} (launches independent child runs when a reaction must be decomposed), \texttt{calculate\_neb\_barrier\_height} (computes segment barrier heights from NEB energies), and \texttt{compare\_structure\_displacements} (checks whether an optimized intermediate geometry is geometrically distinct from the IS and FS, used to confirm true intermediate stability before spawning child runs).

\subsection{Execution Agent}

\begin{promptbox}[Execution Agent System Prompt]
\small
\textbf{Role.} You are the Execution Agent. Your role is to execute a \texttt{SimulationPlan} by orchestrating a sequence of specialized sub-agents.

\textbf{Per-step procedure (in order, no skipping).}
(1) Compute parallelization parameters via \texttt{calculate\_vasp\_parallel\_params}, set \texttt{NCORE} in the step's INCAR, then invoke \texttt{invoke\_vasp\_inputs\_generator\_agent}.
(2) Submit the VASP job via \texttt{invoke\_job\_submission\_agent}. For two independent GO steps (IS and FS relaxation) preceding NEB, submit both in the same invocation before pausing. After any submission, return \texttt{PAUSED\_WAITING\_FOR\_JOBS}.
(3) On restart, parse outputs via \texttt{invoke\_vasp\_outputs\_analyzer\_agent}. For NEB steps, if the parser reports suspicious signatures (\texttt{NEB\_INTERMEDIATE\_MINIMUM}, \texttt{NEB\_FORCE\_PLATEAU\_OR\_OSCILLATION}, \texttt{NEB\_EXPLODING\_IMAGES}, \texttt{NEB\_NSW\_STOP\_CONVERGING}) or if any internal image force exceeds 0.20\,eV/\AA, invoke \texttt{invoke\_neb\_visual\_summary\_agent} and merge its output into the step status before proceeding.

\textbf{Failure handling.} If the parser reports failure for a required step, call \texttt{build\_failure\_event\_from\_parser\_output} to construct a structured \texttt{FailureEvent} and return \texttt{FAILED\_AT\_STEP}. If an intermediate minimum is detected (\texttt{NEB\_INTERMEDIATE\_MINIMUM}), treat it as a replan trigger regardless of convergence and return \texttt{REPLAN\_REQUESTED}. Update \texttt{all\_plan\_step\_statuses} after every step analysis so that the Planning Agent receives full diagnostic context on the next invocation.

\textbf{Output.} Return an \texttt{ExecutionAgentOutput} with status \texttt{PAUSED\_WAITING\_FOR\_JOBS}, \texttt{FAILED\_AT\_STEP}, \texttt{REPLAN\_REQUESTED}, or \texttt{ALL\_STEPS\_SUCCESS}.
\end{promptbox}

The EA's tool registry comprises: \texttt{invoke\_vasp\_inputs\_generator\_agent}, \texttt{invoke\_job\_submission\_agent}, \texttt{invoke\_vasp\_outputs\_analyzer\_agent}, \texttt{invoke\_neb\_visual\_summary\_agent}, \texttt{build\_failure\_event\_from\_parser\_output}, and \texttt{calculate\_vasp\_parallel\_params}. Each \texttt{invoke\_*} call dispatches to a separate LLM-backed agent with its own session.

The EA invokes the Visual Analyzer Agent when any of the following conditions holds after NEB output parsing: the parser status is \texttt{FAILURE}; one or more trigger signatures (\texttt{NEB\_INTERMEDIATE\_MINIMUM}, \texttt{NEB\_FORCE\_PLATEAU\_OR\_OSCILLATION}, \texttt{NEB\_EXPLODING\_IMAGES}, \texttt{NEB\_NSW\_STOP\_CONVERGING}) appear in the failure or warning candidates; or the maximum per-image force among internal images exceeds 0.20\,eV/\AA. This threshold captures cases where convergence has been reached nominally but the path geometry may still be problematic.

\subsection{Inputs Generator Agent}

\begin{promptbox}[VASP Inputs Generator Agent System Prompt]
\small
You are the VASP Input Generator Agent, responsible for preparing the four VASP input files (INCAR, POSCAR, POTCAR, KPOINTS) required for a DFT calculation. Given a \texttt{SimulationStep}, select the appropriate tool: \texttt{generate\_geometry\_optimization\_step\_inputs}, \texttt{generate\_NEB\_step\_inputs}, or \texttt{generate\_VFA\_step\_inputs} for fresh calculations; \texttt{prepare\_restart\_step\_inputs} for GO or NEB steps where \texttt{restart\_from} is set. For NEB inputs, generate interpolated images using the IDPP method and write per-image POSCAR files to numbered subdirectories (\texttt{00/}, \texttt{01/}, \ldots). Do not modify any field of the provided \texttt{SimulationStep}.
\end{promptbox}

The Inputs Generator Agent maps \texttt{SimulationStep} fields to VASP input files as follows. INCAR tags are written directly from the typed \texttt{step\_INCAR} object (dispatching to \texttt{GO\_INCAR}, \texttt{NEB\_INCAR}, or \texttt{VFA\_INCAR}), with Python booleans converted to \texttt{.TRUE.}/\texttt{.FALSE.}. The POSCAR is produced by reading the input geometry with ASE, sorting atoms by chemical symbol for POTCAR consistency, and writing in VASP5 direct-coordinate format. For NEB, IDPP interpolation generates the internal images and spectator atoms (those displaced by less than \texttt{move\_threshold} between IS and FS) are aligned back to IS positions before interpolation to prevent spurious long-range displacements. POTCAR is assembled from the cluster's PAW-PBE library. KPOINTS uses a Gamma-centered Monkhorst-Pack grid. The dipole correction center (\texttt{DIPOL}) is set to the fractional coordinates of the system's center of mass.

\subsection{Job Submission Agent}

\begin{promptbox}[Job Submission Agent System Prompt]
\small
You are the Job Submission Agent. Your task is to submit a VASP job to SLURM. Invoke the pre-configured submission script at \texttt{scripts/submit\_vasp\_simulation\_job\_v2.sh} with the step execution directory, job name, and number of tasks per node. Parse the standard output for the SLURM job ID (format: \texttt{SLURM job ID: \$SLURM\_JOB\_ID}) and return it in the output.
\end{promptbox}

The submission script is a self-submitting SBATCH script: when called on the login node it invokes \texttt{sbatch} on itself, overriding the \texttt{--job-name}, \texttt{--ntasks-per-node}, \texttt{--time}, and \texttt{--account} headers at submission time. The wall-clock limit and SLURM account are drawn from environment variables (\texttt{AGENTIC\_TS\_SLURM\_TIME}, \texttt{AGENTIC\_TS\_SLURM\_ACCOUNT}), allowing overrides without modifying the script. On the compute node, VASP is executed in the step execution directory with per-job stdout and stderr captured to a timestamped \texttt{slurm/} subdirectory.

\subsection{Outputs Parser Agent}

\begin{promptbox}[VASP Outputs Parser Agent System Prompt]
\small
You are the VASP Outputs Parser Agent. Given a \texttt{SimulationStep} and its execution directory, parse the VASP output files and return a structured \texttt{VASPOutputsParserAgentOutput}. Use \texttt{parse\_geometry\_optimization\_outputs}, \texttt{parse\_NEB\_outputs}, or \texttt{parse\_vibrational\_frequency\_analysis\_outputs} according to the step type. Do not inspect raw VASP logs directly; use only the provided parsing tools. Do not invoke the NEB visual debugger; return parser diagnostics and artifact paths only. If \texttt{job\_id} is unavailable, proceed from \texttt{step\_execution\_directory} without blocking.
\end{promptbox}

The parser agent's tools operate on VASP's \texttt{OUTCAR} and \texttt{OSZICAR} files using pattern-based extractors rather than a general log parser. This design avoids hallucination that can occur when an LLM reads raw, megabyte-scale output files: the parser tools return structured numeric arrays that are safe to embed in the LLM context.

The principal extractors and the quantities they compute are summarized below.

\medskip
\noindent\textit{Geometry Optimization:}
\begin{itemize}[leftmargin=0.5cm]
  \item \texttt{check\_completion} / \texttt{check\_convergence}: detect VASP's \texttt{reached required accuracy} marker in the tail of OUTCAR.
  \item \texttt{parse\_go\_ionic\_force\_history}: regex over \texttt{FORCES acting on ions} blocks; returns the max-force series over the last 10 ionic steps.
  \item \texttt{parse\_go\_energy\_sigma0\_history}: extracts \texttt{energy(sigma->0)} from OUTCAR energy blocks.
  \item \texttt{parse\_go\_scf\_iterations\_history}: counts SCF iterations per ionic step from OSZICAR \texttt{N} records.
  \item \texttt{detect\_scf\_non\_convergence} / \texttt{detect\_ionic\_non\_convergence}: flag convergence failures from OUTCAR warning strings.
\end{itemize}

\noindent\textit{Nudged Elastic Band:}
\begin{itemize}[leftmargin=0.5cm]
  \item \texttt{parse\_NEB\_OUTCARs}: reads per-image OUTCAR files in numbered subdirectories; extracts final energies ($\sigma\!\to\!0$) and max forces per image; identifies the highest-energy internal image as the candidate TS.
  \item \texttt{parse\_neb\_per\_image\_force\_histories}: returns force-vs-step history (last 10 steps) for each image, enabling plateau/oscillation detection.
  \item \texttt{detect\_neb\_force\_plateau\_or\_oscillation}: checks whether the final per-image force variance falls below a tolerance while forces remain above EDIFFG.
  \item \texttt{detect\_neb\_exploding\_images}: flags any image with final force exceeding 1.5\,eV/\AA\ as unphysical.
  \item \texttt{detect\_neb\_intermediate\_minimum}: identifies an interior energy minimum satisfying both an energy depth threshold and per-image force criterion.
  \item \texttt{visualize\_NEB\_outputs}: generates per-image \texttt{CONTCAR.png} previews, POSCAR/OUTCAR trajectory GIFs, energy-vs-image and force-vs-image plots, which are attached as artifact paths.
\end{itemize}

\noindent\textit{Vibrational Frequency Analysis:}
\begin{itemize}[leftmargin=0.5cm]
  \item \texttt{parse\_VFA\_OUTCAR}: extracts the full vibrational mode spectrum from OUTCAR \texttt{THz} blocks; separates imaginary and real modes.
  \item \texttt{resolve\_vfa\_imag\_threshold}: reads \texttt{imaginary\_frequency\_threshold} from the step schema (default $-10$\,meV) to filter numerical noise modes.
  \item \texttt{compute\_vfa\_mode\_count\_metrics}: counts modes above threshold; sets \texttt{first\_order\_saddle\_point = True} when exactly one imaginary mode exceeds threshold.
\end{itemize}

The pattern-match-first design means the parser agent never needs to reason about file content; its only task is to route each step type to the correct tool and return the structured output. This keeps the LLM's role in output analysis to tool selection and result interpretation, not text parsing, which substantially reduces error rates on large OUTCAR files.

\subsection{Visual Analyzer Agent}
\label{app:visual-analyzer}

The NEB Visual Summary Agent is a multimodal debugger that examines rendered pathway images to identify chemistry and geometry events that numeric diagnostics alone cannot detect. It is invoked by the Execution Agent as a tool call (\texttt{invoke\_neb\_visual\_summary\_agent}) and uses \texttt{gpt-5.4} with \texttt{reasoning.effort = medium}.

\medskip
\noindent\textbf{Rendering pipeline.} Before invoking the agent, the orchestrator renders a pathway montage from the NEB image directories. Each row of the montage corresponds to one NEB image; each row contains three 2D projections of the same 3D structure: XY (top view, \texttt{0x,0y,0z}), XZ (front view, \texttt{90x,0y,0z}), and YZ (side view, \texttt{0x,90y,0z}). Atom colors follow ASE element conventions; a color legend image is appended to the visual context. Alongside the montage, the agent receives the energy-vs-image and force-vs-image plots generated by the Outputs Parser Agent, and a JSON context block containing the parser's numeric diagnostics, candidate TS image index, and failure signature candidates.

\begin{promptbox}[NEB Visual Summary Agent System Prompt]
\small
You are the NEB Visual Summary Agent, a multimodal debugger for Nudged Elastic Band pathways. You will receive a montage of NEB images (one row per image, three orthogonal 2D projections per row), energy and force plots, and an ASE color legend.

Treat each row as one 3D structure shown from three angles. Use agreement across projections before concluding that a bond breaks, a fragment rotates, desorbs, or collides. If the evidence is ambiguous, emit \texttt{UNCERTAIN} rather than guessing. This is a debugging task, not a captioning task: focus on chemistry-relevant events (bond breaking/forming, rotation, translation, desorption, surface reconstruction, intermediate stabilization, atom overlap).

Correlate the montage with the energy and force plots: an interior energy peak is the candidate TS region; high forces on specific images indicate incomplete convergence; very high forces with geometric distortions argue against restarting from the current geometry; flat or oscillatory forces without geometric progress suggest optimizer plateau.

Return a structured \texttt{NEBVisualSummaryAgentOutput}.
\end{promptbox}

The visual user prompt (\texttt{neb\_visual\_summary\_user\_prompt.md}) is rendered at call time by substituting a JSON context block containing parser diagnostics and per-image numeric values into a template. The agent has no tools; its only output is the structured schema below.

\begin{lstlisting}[language=Python, basicstyle=\small\ttfamily,
    commentstyle=\color{gray}\itshape, columns=fullflexible]
class NEBVisualSummary:
    status: Literal["SUCCESS", "UNAVAILABLE"]  # UNAVAILABLE if artifacts missing
    overall_summary: str           # High-level pathway interpretation
    pathway_character: Literal[    # Coarse plausibility verdict
        "PHYSICAL_PATHWAY", "LIKELY_UNPHYSICAL", "AMBIGUOUS"
    ]
    key_events: List[NEBVisualEvent]  # Ordered list of localized events (see below)
    recommended_debug_focus: List[str]  # Actionable debugging priorities
    unphysicality_signals: List[str]    # Specific reasons pathway looks unphysical
    artifact_paths: Dict[str, str]      # Paths to montage, manifest, plots
    limitations: Optional[str]          # Ambiguity caveats; notes missing artifacts

class NEBVisualEvent:
    event_type: Literal[           # Chemistry/geometry label
        "BOND_BREAKING", "BOND_FORMING", "ROTATION", "TRANSLATION",
        "DESORPTION", "SURFACE_RECONSTRUCTION",
        "INTERMEDIATE_STABILIZATION", "ATOM_OVERLAP_OR_COLLISION", "UNCERTAIN"
    ]
    start_image_idx: int           # Inclusive start of the event
    end_image_idx: int             # Inclusive end of the event
    confidence: float              # Model confidence in [0, 1]
    atoms_involved: List[str]      # Element labels, e.g. ["O", "H", "Pt surface"]
    summary: str                   # Short description
    debug_implication: str         # How this event should influence replanning
\end{lstlisting}

The structured output is merged back into the \texttt{NudgedElasticBandOutput} diagnostics by the \texttt{merge\_visual\_summary\_into\_parser\_output} function before the Execution Agent constructs the \texttt{FailureEvent}, ensuring that the Planning Agent receives both numeric and visual evidence in the same structured payload.

\section{Expert-Crafted Debugging Guidelines}
\label{app:debugging-guidelines}

The Planning Agent's replan prompt includes a set of debugging guidelines loaded from \texttt{prompts/debugging\_guidelines.md}. These guidelines were written by the research team based on accumulated experience running VASP NEB calculations and encode interventions that are difficult to derive from first principles alone (e.g.\ when to crop a path, the MAXMOVE heuristic for near-threshold NEB runs, or the energy tolerance for confirming a stable intermediate). They are provided as injected context rather than hardcoded logic so that the Planning Agent can combine them with its own domain reasoning and override them when the evidence warrants it. Below is the full content as injected into the replan prompt.

\begin{promptbox}[Expert-Crafted Debugging Guidelines]
\small
\textbf{Infrastructure failures.} If the SLURM job status indicates a node failure or preemption, continue with the current plan if the calculation would likely have converged absent the failure; otherwise generate a revised plan.

\medskip
\textbf{Geometry Optimization failures.}
(1) If max forces are large and not converging, reduce POTIM. Apply this intervention at most once.
(2) If reducing POTIM does not resolve the issue, switch to a two-phase strategy: a low-cost pre-relaxation (ENCUT 250\,eV, k-mesh 2$\times$2$\times$1, EDIFFG $-$0.2\,eV/\AA) followed by a final GO at target settings initialized from the pre-relaxed geometry.

\medskip
\textbf{NEB failures.}
Apply at most one or a few closely related interventions at a time.

Case 1: Forces are only slightly above the EDIFFG threshold and steadily decreasing. Restart for additional steps (increase NSW) without other changes.

Case 2: Forces are significantly above threshold, or the job timed out before reaching NSW.

\textit{Fix 2a}: If all per-image forces are close to EDIFFG, restart from the current point with a lower MAXMOVE. For coarse NEB, also consider disabling climbing image.

\textit{Fix 2b}: If the energy profile shows a clear interior maximum and forces at and around that image are below 0.2\,eV/\AA, crop the path to the images nearest the maximum. Apply cropping at most once per IS/FS pair; check the replan history before applying.

Note: If final forces are unphysically large ($>$1.5\,eV/\AA) after many ionic steps, do not restart from the current geometry. Apply other fixes or escalate.

Case 3: Intermediate minimum (\texttt{NEB\_INTERMEDIATE\_MINIMUM}). Before any action, verify that the parent reaction has not already been split (each parent may spawn at most two child runs; child runs cannot spawn further). Run a GO on the intermediate image and compare the optimized energy and structure to IS and FS using \texttt{compare\_structure\_displacements}. If the GO energy changes by less than 0.2\,eV and the structure remains distinct from both IS and FS, use \texttt{spawn\_multi\_step\_ts\_search} to launch IS$\to$INT and INT$\to$FS child runs; skip a segment if its barrier is below 0.1\,eV. Escalate with child run IDs once spawned.

Case 4: Monotonically increasing or decreasing energy profile with forces converged below 0.2\,eV/\AA\ indicates a barrierless reaction. Escalate to the user.

\medskip
\textbf{Vibrational Frequency Analysis failures.}
Ignore imaginary modes with $|$frequency$|$ below 10\,cm$^{-1}$ as numerical noise.

(1) Zero or more than one imaginary mode after converged NEB: crop around the highest-energy NEB image and re-run VFA.
(2) If cropping does not help: tighten VFA convergence (lower EDIFF, reduce POTIM to 0.01\,\AA). Apply at most once.
(3) If still unsuccessful: tighten the preceding NEB convergence (lower EDIFFG and EDIFF). Apply at most once.
(4) If none of the above resolves the issue: escalate; the candidate TS is likely not a first-order saddle point.

\medskip
\textbf{Restarting from an existing point.} Provide a new \texttt{step\_execution\_directory} and set \texttt{restart\_from} to the previous step's directory. Do not set \texttt{restart\_from} equal to \texttt{step\_execution\_directory}.

\medskip
\textbf{General guardrails.} Do not use NCORE as a replanning parameter (set at runtime). Do not repeat debugging interventions that appear in the replan log for the same evidence. Ensure that the final NEB ENCUT matches the GO ENCUT at all times, even when only one is being replanned.
\end{promptbox}

These guidelines were developed iteratively by the research team over a series of trial runs and draw on community-established best practices for VASP NEB calculations ~\cite{henkelman-ci-neb}. They are not exhaustive; the Planning Agent is explicitly instructed to apply domain reasoning beyond the guidelines when it encounters failure modes not covered there.

\section{OC20NEB Simulation Details and Evaluation Cases}
\label{app:oc20neb-cases}

\subsection{DFT Settings for OC20NEB Analysis}
All OC20NEB calculations were performed with VASP6.3 using projector-augmented-wave potentials and a plane-wave basis set \cite{kresse_efficiency_1996, kresse_efficient_1996, kresse_ultrasoft_1999}. Exchange and correlation were treated with the revised Perdew--Burke--Ernzerhof functional\cite{hammer_improved_1999}, and dispersion was included via Grimme's D3 correction with Becke--Johnson damping \cite{grimme_consistent_2010, grimme_effect_2011}. The Brillouin zone was sampled with a $2\times2\times1$ \textit{k}-point grid, and Gaussian smearing with a width of $0.05~\mathrm{eV}$ was applied to the orbital occupancies. Calculations were non-spin-polarized and performed without symmetry. For the surface slab geometries, only atomic positions were relaxed; the cell shape and volume were held fixed (\texttt{ISIF=2}), and a dipole correction to the potential and forces was applied along the slab-normal direction. A plane wave kinetic-energy cutoff of $350~\mathrm{eV}$  and a force-based stopping criterion of $0.05~\mathrm{eV}\,\AA^{-1}$ was used for all convergence of initial and final geometry optimizations and climbing image NEB calculations. Candidate transition-state images from the NEB were validated by finite-difference vibrational frequency analysis (VFA) cutoff and $2\times2\times1$ \textit{k}-point grid. The agent was permitted to modify these computational parameters during replanning; the values reported above correspond to the final settings used for all calculations included in this section.

\subsection{Compute Budget for OC20NEB Cases}
\label{oc20neb_tsa_config}

A maximum of five replanning attempts was permitted per case (i.e., one initial plan followed by up to five revisions), after which the case was terminated as unresolved. During every replan for each VASP job (Geometry Optimization, Nudge Elastic Band calculation, and Vibration Frequency Analysis), a computational budget of 2048 cpu-hours was allocated.

\edef\savedTabColSep{\the\tabcolsep}%
\edef\savedArrayStretch{\arraystretch}%
\setlength{\tabcolsep}{4pt}%
\renewcommand{\arraystretch}{1.12}%
\setlength{\LTleft}{\fill}%
\setlength{\LTright}{\fill}%
\scriptsize
\begin{longtable}{%
  >{\raggedright\arraybackslash}p{0.22\linewidth}
  >{\raggedright\arraybackslash}p{0.15\linewidth}
  >{\raggedright\arraybackslash}p{0.17\linewidth}
  >{\raggedright\arraybackslash}p{0.08\linewidth}
  >{\raggedright\arraybackslash}p{0.26\linewidth}}
\captionsetup{
  font=normalsize,          
  justification=raggedright,
  singlelinecheck=false
}
\caption{All OC20NEB evaluation cases used in the 100-case benchmark. The table spans diverse dissociation and H-transfer reactions across a broad range of catalyst surfaces.
  \textit{Surface} gives the reduced slab formula;
  \textit{Ads.}\ gives the Hill-style adsorbate formula;
  \textit{Reaction} reports bond-length changes between the first and last NEB images.}
\label{tab:oc20neb_evaluation_all_cases} \\

\toprule
\textbf{Case} & \textbf{class / facet} & \textbf{Surface} &
\textbf{Ads.} & \textbf{Reaction and bond change} \\
\midrule
\endfirsthead

\multicolumn{5}{l}{\small\itshape (Table~\ref{tab:oc20neb_evaluation_all_cases} continued)} \\[2pt]
\toprule
\textbf{Case} & \textbf{Split / class / facet} & \textbf{Surface} &
\textbf{Ads.} & \textbf{Reaction and bond change} \\
\midrule
\endhead

\midrule
\multicolumn{5}{r}{\small\itshape continued on next page} \\
\endfoot

\bottomrule
\endlastfoot

\makecell[tl]{\texttt{dissociation\_id\_62\_}\\\texttt{7898\_44\_111-0\_neb1.0}} & \makecell[tl]{dissociation\\(111), site~0} & PTe$_{2}$Ti$_{2}$ & C$_{2}$H$_{3}$O & Broken: C--C: 1.348 to 3.090~\AA. \\
\makecell[tl]{\texttt{dissociation\_id\_63\_}\\\texttt{6310\_46\_011-0\_neb1.0}} & \makecell[tl]{dissociation\\(011), site~0} & Fe$_{2}$ScSi$_{2}$ & HN & Broken: H--N: 1.035 to 2.749~\AA. \\
\makecell[tl]{\texttt{dissociation\_id\_75\_}\\\texttt{2211\_45\_111-0\_neb1.0}} & \makecell[tl]{dissociation\\(111), site~0} & Ge$_{2}$Hf$_{3}$ & N$_{2}$ & Broken: N--N: 1.336 to 2.883~\AA. \\
\makecell[tl]{\texttt{dissociation\_id\_79\_}\\\texttt{5183\_39\_011-8\_neb1.0}} & \makecell[tl]{dissociation\\(011), site~8} & Ca$_{3}$Ga$_{2}$N$_{4}$ & C$_{2}$H$_{2}$O$_{2}$ & Broken: C--C: 1.566 to 2.825~\AA. \\
\makecell[tl]{\texttt{dissociation\_id\_81\_}\\\texttt{1944\_46\_222-1\_neb1.0}} & \makecell[tl]{dissociation\\(222), site~1} & HgTi$_{3}$ & HN & Broken: H--N: 1.040 to 2.613~\AA. \\
\makecell[tl]{\texttt{dissociation\_id\_105\_}\\\texttt{742\_53\_111-4\_neb1.0}} & \makecell[tl]{dissociation\\(111), site~4} & Cu$_{3}$Ge & H$_{2}$ & Broken: H--H: 0.773 to 2.652~\AA. \\
\makecell[tl]{\texttt{dissociation\_id\_108\_}\\\texttt{99\_21\_111-0\_neb1.0}} & \makecell[tl]{dissociation\\(111), site~0} & Ca & C$_{2}$HO & Broken: C--H: 1.148 to 3.038~\AA. \\
\makecell[tl]{\texttt{dissociation\_id\_114\_}\\\texttt{11171\_51\_200-3\_neb1.0}} & \makecell[tl]{dissociation\\(200), site~3} & Bi$_{2}$Ni$_{3}$S$_{2}$ & H$_{3}$N & Broken: H--N: 1.028 to 2.547~\AA. \\
\makecell[tl]{\texttt{dissociation\_id\_127\_}\\\texttt{9635\_0\_100-1\_neb1.0}} & \makecell[tl]{dissociation\\(100), site~1} & GaTcTi$_{2}$ & HO & Broken: H--O: 0.977 to 3.001~\AA. \\
\makecell[tl]{\texttt{dissociation\_id\_132\_}\\\texttt{4388\_18\_111-0\_neb1.0}} & \makecell[tl]{dissociation\\(111), site~0} & CoIrTi$_{2}$ & C$_{2}$O & Broken: C--C: 1.320 to 4.259~\AA. \\
\makecell[tl]{\texttt{dissociation\_id\_141\_}\\\texttt{476\_27\_100-3\_neb1.0}} & \makecell[tl]{dissociation\\(100), site~3} & Ir$_{3}$W & C$_{2}$H$_{2}$O & Broken: C--C: 1.377 to 3.181~\AA. \\
\makecell[tl]{\texttt{dissociation\_id\_143\_}\\\texttt{7685\_39\_200-0\_neb1.0}} & \makecell[tl]{dissociation\\(200), site~0} & KMnTe$_{2}$ & C$_{2}$H$_{2}$O$_{2}$ & Broken: C--C: 1.403 to 4.051~\AA. \\
\makecell[tl]{\texttt{dissociation\_id\_154\_}\\\texttt{1581\_1\_111-0\_neb1.0}} & \makecell[tl]{dissociation\\(111), site~0} & GaPt$_{3}$ & CO & Broken: C--O: 1.169 to 3.705~\AA. \\
\makecell[tl]{\texttt{dissociation\_id\_171\_}\\\texttt{8597\_46\_222-4\_neb1.0}} & \makecell[tl]{dissociation\\(222), site~4} & IrPS & HN & Broken: H--N: 1.027 to 2.656~\AA. \\
\makecell[tl]{\texttt{dissociation\_id\_175\_}\\\texttt{9905\_25\_111-3\_neb1.0}} & \makecell[tl]{dissociation\\(111), site~3} & SSeSn & C$_{2}$HO$_{2}$ & Broken: C--O: 1.221 to 3.150~\AA. \\
\makecell[tl]{\texttt{dissociation\_id\_181\_}\\\texttt{116\_6\_111-0\_neb1.0}} & \makecell[tl]{dissociation\\(111), site~0} & Zn & CHO & Broken: C--O: 1.257 to 3.870~\AA. \\
\makecell[tl]{\texttt{dissociation\_id\_182\_}\\\texttt{2807\_43\_111-2\_neb1.0}} & \makecell[tl]{dissociation\\(111), site~2} & Sc$_{5}$Si$_{3}$ & C$_{2}$H$_{3}$O & Broken: C--C: 1.539 to 3.295~\AA. \\
\makecell[tl]{\texttt{dissociation\_id\_183\_}\\\texttt{7231\_38\_211-0\_neb1.0}} & \makecell[tl]{dissociation\\(211), site~0} & AgOsSc$_{2}$ & C$_{2}$H$_{2}$O$_{2}$ & Broken: C--C: 1.366 to 3.518~\AA. \\
\makecell[tl]{\texttt{dissociation\_id\_211\_}\\\texttt{4455\_16\_111-0\_neb1.0}} & \makecell[tl]{dissociation\\(111), site~0} & Co$_{2}$GeTi & CH$_{4}$O & Broken: H--O: 0.984 to 2.851~\AA. \\
\makecell[tl]{\texttt{dissociation\_id\_219\_}\\\texttt{10897\_5\_100-0\_neb1.0}} & \makecell[tl]{dissociation\\(100), site~0} & Ga$_{3}$NbZn$_{3}$ & CHO & Broken: C--H: 1.108 to 2.828~\AA. \\
\makecell[tl]{\texttt{dissociation\_id\_228\_}\\\texttt{10384\_22\_211-3\_neb1.0}} & \makecell[tl]{dissociation\\(211), site~3} & RhSbTe & C$_{2}$HO & Broken: C--O: 1.204 to 2.903~\AA. \\
\makecell[tl]{\texttt{dissociation\_id\_234\_}\\\texttt{6939\_25\_111-3\_neb1.0}} & \makecell[tl]{dissociation\\(111), site~3} & Ga$_{2}$RuSc & C$_{2}$HO$_{2}$ & Broken: C--O: 1.367 to 2.930~\AA. \\
\makecell[tl]{\texttt{dissociation\_id\_256\_}\\\texttt{4371\_5\_222-0\_neb1.0}} & \makecell[tl]{dissociation\\(222), site~0} & CuSb$_{3}$Ti$_{2}$ & CHO & Broken: C--H: 1.106 to 2.500~\AA. \\
\makecell[tl]{\texttt{dissociation\_id\_266\_}\\\texttt{9844\_9\_111-2\_neb1.0}} & \makecell[tl]{dissociation\\(111), site~2} & Ge$_{6}$PdY$_{2}$ & CH$_{2}$O & Broken: C--O: 1.446 to 2.962~\AA. \\
\makecell[tl]{\texttt{dissociation\_id\_271\_}\\\texttt{7189\_22\_211-5\_neb1.0}} & \makecell[tl]{dissociation\\(211), site~5} & CdPtSr & C$_{2}$HO & Broken: C--O: 1.221 to 4.006~\AA. \\
\makecell[tl]{\texttt{dissociation\_id\_283\_}\\\texttt{4512\_22\_122-3\_neb1.0}} & \makecell[tl]{dissociation\\(122), site~3} & SbSnTi & C$_{2}$HO & Broken: C--O: 1.329 to 3.333~\AA. \\
\makecell[tl]{\texttt{dissociation\_id\_287\_}\\\texttt{5912\_11\_100-1\_neb1.0}} & \makecell[tl]{dissociation\\(100), site~1} & CuPd$_{2}$Sn & CH$_{2}$O & Broken: H--O: 0.983 to 3.124~\AA. \\
\makecell[tl]{\texttt{dissociation\_id\_294\_}\\\texttt{2408\_32\_222-0\_neb1.0}} & \makecell[tl]{dissociation\\(222), site~0} & AlSc$_{3}$ & C$_{2}$H$_{2}$O & Broken: C--C: 1.386 to 3.103~\AA. \\
\makecell[tl]{\texttt{dissociation\_id\_359\_}\\\texttt{10123\_21\_211-2\_neb1.0}} & \makecell[tl]{dissociation\\(211), site~2} & FeNiPt$_{6}$ & C$_{2}$HO & Broken: C--H: 1.108 to 2.962~\AA. \\
\makecell[tl]{\texttt{dissociation\_id\_425\_}\\\texttt{7980\_18\_222-0\_neb1.0}} & \makecell[tl]{dissociation\\(222), site~0} & AlRh$_{2}$Sc & C$_{2}$O & Broken: C--C: 1.316 to 3.124~\AA. \\
\makecell[tl]{\texttt{dissociation\_id\_569\_}\\\texttt{2986\_44\_111-0\_neb1.0}} & \makecell[tl]{dissociation\\(111), site~0} & Cu$_{5}$Zr & C$_{2}$H$_{3}$O & Broken: C--C: 1.345 to 4.157~\AA. \\
\makecell[tl]{\texttt{dissociation\_id\_631\_}\\\texttt{8786\_1\_111-0\_neb1.0}} & \makecell[tl]{dissociation\\(111), site~0} & GaOs$_{2}$V & CO & Broken: C--O: 1.180 to 3.107~\AA. \\
\makecell[tl]{\texttt{dissociation\_id\_652\_}\\\texttt{8243\_0\_111-9\_neb1.0}} & \makecell[tl]{dissociation\\(111), site~9} & AuTiZr & HO & Broken: H--O: 0.975 to 2.498~\AA. \\
\makecell[tl]{\texttt{dissociation\_id\_657\_}\\\texttt{11403\_24\_222-1\_neb1.0}} & \makecell[tl]{dissociation\\(222), site~1} & Bi$_{3}$CuZr$_{5}$ & C$_{2}$HO & Broken: C--O: 1.370 to 3.553~\AA. \\
\makecell[tl]{\texttt{dissociation\_ood\_70\_}\\\texttt{8416\_46\_111-1\_neb1.0}} & \makecell[tl]{dissociation\\(111), site~1} & IrSnTi & HN & Broken: H--N: 1.038 to 3.305~\AA. \\
\makecell[tl]{\texttt{dissociation\_ood\_82\_}\\\texttt{7708\_18\_211-5\_neb1.0}} & \makecell[tl]{dissociation\\(211), site~5} & FeTe$_{2}$Zr$_{6}$ & C$_{2}$O & Broken: C--C: 1.360 to 3.709~\AA. \\
\makecell[tl]{\texttt{dissociation\_ood\_85\_}\\\texttt{7443\_10\_211-1\_neb1.0}} & \makecell[tl]{dissociation\\(211), site~1} & Rh$_{2}$SbSc & CH$_{2}$O & Broken: C--H: 1.107 to 2.488~\AA. \\
\makecell[tl]{\texttt{dissociation\_ood\_153\_}\\\texttt{6035\_49\_211-0\_neb1.0}} & \makecell[tl]{dissociation\\(211), site~0} & Ag$_{2}$Si$_{2}$Sr & HN$_{2}$O & Broken: N--N: 1.294 to 4.002~\AA. \\
\makecell[tl]{\texttt{dissociation\_ood\_295\_}\\\texttt{10297\_43\_211-1\_neb1.0}} & \makecell[tl]{dissociation\\(211), site~1} & CuHf$_{2}$Re & C$_{2}$H$_{3}$O & Broken: C--C: 1.524 to 3.658~\AA. \\
\makecell[tl]{\texttt{dissociation\_ood\_350\_}\\\texttt{7417\_25\_111-4\_neb1.0}} & \makecell[tl]{dissociation\\(111), site~4} & PRuSe & C$_{2}$HO$_{2}$ & Broken: C--O: 1.355 to 2.670~\AA. \\
\makecell[tl]{\texttt{dissociation\_ood\_378\_}\\\texttt{9169\_48\_211-0\_neb1.0}} & \makecell[tl]{dissociation\\(211), site~0} & CdPbRh$_{2}$ & CN & Broken: C--N: 1.276 to 3.088~\AA. \\
\makecell[tl]{\texttt{dissociation\_ood\_457\_}\\\texttt{10516\_11\_211-0\_neb1.0}} & \makecell[tl]{dissociation\\(211), site~0} & CoN$_{2}$Sr$_{2}$ & CH$_{2}$O & Broken: H--O: 0.978 to 3.020~\AA. \\
\makecell[tl]{\texttt{dissociation\_ood\_523\_}\\\texttt{6324\_2\_000-1\_neb1.0}} & \makecell[tl]{dissociation\\(000), site~1} & CuGeSc & CH & Broken: C--H: 1.105 to 2.850~\AA. \\
\makecell[tl]{\texttt{dissociation\_ood\_528\_}\\\texttt{7131\_18\_111-1\_neb1.0}} & \makecell[tl]{dissociation\\(111), site~1} & Co$_{2}$Ge$_{3}$Sc$_{3}$ & C$_{2}$O & Broken: C--C: 1.321 to 3.133~\AA. \\
\makecell[tl]{\texttt{dissociation\_ood\_542\_}\\\texttt{6869\_21\_111-4\_neb1.0}} & \makecell[tl]{dissociation\\(111), site~4} & Au$_{2}$Sc$_{2}$Sn & C$_{2}$HO & Broken: C--H: 1.085 to 2.660~\AA. \\
\makecell[tl]{\texttt{dissociation\_ood\_664\_}\\\texttt{6163\_1\_111-1\_neb1.0}} & \makecell[tl]{dissociation\\(111), site~1} & OsPZr & CO & Broken: C--O: 1.299 to 3.001~\AA. \\
\makecell[tl]{\texttt{transfer\_id\_1\_}\\\texttt{4776\_2\_211-0\_neb1.0}} & \makecell[tl]{H transfer\\(211), site~0} & NiSnTi & C$_{2}$H$_{2}$ & Broken: C--H: 1.150 to 3.281~\AA. Formed: C--H: 3.395 to 1.108~\AA. \\
\makecell[tl]{\texttt{transfer\_id\_2\_}\\\texttt{10414\_20\_122-4\_neb1.0}} & \makecell[tl]{H transfer\\(122), site~4} & PtRb$_{2}$S$_{2}$ & CH$_{4}$O & Broken: C--O: 1.450 to 7.030~\AA. Formed: H--O: 3.044 to 0.988~\AA. \\
\makecell[tl]{\texttt{transfer\_id\_9\_}\\\texttt{10432\_4\_122-5\_neb1.0}} & \makecell[tl]{H transfer\\(122), site~5} & GeNiZr & CH$_{3}$O & Broken: H--O: 0.979 to 2.657~\AA. Formed: C--H: 2.925 to 1.112~\AA. \\
\makecell[tl]{\texttt{transfer\_id\_12\_}\\\texttt{6537\_3\_211-1\_neb1.0}} & \makecell[tl]{H transfer\\(211), site~1} & AlRu$_{2}$Zr & C$_{2}$H$_{4}$ & Broken: C--H: 1.106 to 4.414~\AA. Formed: C--H: 3.014 to 1.147~\AA. \\
\makecell[tl]{\texttt{transfer\_id\_14\_}\\\texttt{8952\_19\_100-0\_neb1.0}} & \makecell[tl]{H transfer\\(100), site~0} & AgGaY$_{2}$ & H$_{2}$N$_{3}$ & Broken: N--N: 1.466 to 4.336~\AA. Formed: N--N: 3.408 to 1.434~\AA. \\
\makecell[tl]{\texttt{transfer\_id\_25\_}\\\texttt{2929\_0\_111-3\_neb1.0}} & \makecell[tl]{H transfer\\(111), site~3} & HfSn & CH$_{3}$O & Broken: H--O: 0.976 to 3.010~\AA. Formed: C--H: 3.184 to 1.100~\AA. \\
\makecell[tl]{\texttt{transfer\_id\_78\_}\\\texttt{902\_5\_100-1\_neb1.0}} & \makecell[tl]{H transfer\\(100), site~1} & Ge$_{2}$Pt & C$_{2}$H$_{2}$O & Broken: C--H: 1.126 to 3.639~\AA. Formed: C--H: 3.420 to 1.097~\AA. \\
\makecell[tl]{\texttt{transfer\_id\_87\_}\\\texttt{9189\_0\_122-17\_neb1.0}} & \makecell[tl]{H transfer\\(122), site~17} & GeSb$_{7}$Zr$_{4}$ & CH$_{3}$O & Broken: H--O: 0.967 to 2.567~\AA. Formed: C--H: 3.811 to 1.115~\AA. \\
\makecell[tl]{\texttt{transfer\_id\_93\_}\\\texttt{5175\_2\_211-0\_neb1.0}} & \makecell[tl]{H transfer\\(211), site~0} & Ge$_{4}$Na$_{3}$Pt$_{4}$ & C$_{2}$H$_{2}$ & Broken: C--H: 1.105 to 4.238~\AA. Formed: C--H: 4.102 to 1.106~\AA. \\
\makecell[tl]{\texttt{transfer\_id\_102\_}\\\texttt{7779\_7\_211-9\_neb1.0}} & \makecell[tl]{H transfer\\(211), site~9} & CuHf$_{5}$Sb$_{3}$ & CH$_{2}$N & Broken: C--H: 1.143 to 5.336~\AA. Formed: H--N: 4.061 to 1.032~\AA. \\
\makecell[tl]{\texttt{transfer\_id\_118\_}\\\texttt{2583\_2\_211-1\_neb1.0}} & \makecell[tl]{H transfer\\(211), site~1} & Al$_{3}$Sc & C$_{2}$H$_{2}$ & Broken: C--H: 1.116 to 3.994~\AA. Formed: C--H: 3.082 to 1.109~\AA; C--H: 8.497 to 1.114~\AA. \\
\makecell[tl]{\texttt{transfer\_id\_121\_}\\\texttt{2438\_5\_100-0\_neb1.0}} & \makecell[tl]{H transfer\\(100), site~0} & CuZr$_{2}$ & C$_{2}$H$_{2}$O & Broken: C--H: 1.118 to 4.085~\AA. Formed: C--H: 3.447 to 1.121~\AA. \\
\makecell[tl]{\texttt{transfer\_id\_123\_}\\\texttt{6333\_1\_211-4\_neb1.0}} & \makecell[tl]{H transfer\\(211), site~4} & PPt$_{5}$Sn & CH$_{2}$O & Broken: H--O: 0.981 to 3.886~\AA. Formed: C--H: 7.941 to 1.105~\AA. \\
\makecell[tl]{\texttt{transfer\_id\_128\_}\\\texttt{9470\_21\_100-1\_neb1.0}} & \makecell[tl]{H transfer\\(100), site~1} & FeHfSb & C$_{2}$H & Broken: C--C: 1.371 to 3.277~\AA. Formed: C--H: 2.606 to 1.099~\AA. \\
\makecell[tl]{\texttt{transfer\_id\_128\_}\\\texttt{9912\_4\_000-3\_neb1.0}} & \makecell[tl]{H transfer\\(000), site~3} & AsRuSe & CH$_{3}$O & Broken: H--O: 0.984 to 5.052~\AA. Formed: C--H: 3.483 to 1.108~\AA. \\
\makecell[tl]{\texttt{transfer\_id\_129\_}\\\texttt{7600\_1\_211-2\_neb1.0}} & \makecell[tl]{H transfer\\(211), site~2} & MnSiY & CH$_{2}$O & Broken: H--O: 0.974 to 3.658~\AA. \\
\makecell[tl]{\texttt{transfer\_id\_136\_}\\\texttt{3476\_7\_100-0\_neb1.0}} & \makecell[tl]{H transfer\\(100), site~0} & MoTi & CH$_{2}$N & Broken: C--H: 1.106 to 4.062~\AA. Formed: H--N: 2.841 to 1.033~\AA. \\
\makecell[tl]{\texttt{transfer\_id\_140\_}\\\texttt{2720\_7\_100-0\_neb1.0}} & \makecell[tl]{H transfer\\(100), site~0} & Au$_{2}$Sc & CH$_{2}$N & Broken: C--H: 1.111 to 3.979~\AA. Formed: H--N: 3.195 to 1.033~\AA. \\
\makecell[tl]{\texttt{transfer\_id\_170\_}\\\texttt{2959\_23\_2-1-1-1\_neb1.0}} & \makecell[tl]{H transfer\\(2-1-1), site~1} & Si$_{3}$Zr$_{5}$ & C$_{3}$H$_{3}$ & Broken: C--H: 1.104 to 3.486~\AA. Formed: C--H: 3.135 to 1.106~\AA. \\
\makecell[tl]{\texttt{transfer\_id\_173\_}\\\texttt{2519\_21\_100-0\_neb1.0}} & \makecell[tl]{H transfer\\(100), site~0} & Ga$_{3}$Y & C$_{2}$H & Broken: C--C: 1.298 to 3.653~\AA. Formed: C--H: 2.993 to 1.113~\AA. \\
\makecell[tl]{\texttt{transfer\_id\_178\_}\\\texttt{9032\_3\_111-0\_neb1.0}} & \makecell[tl]{H transfer\\(111), site~0} & In$_{3}$Pb$_{3}$Y$_{2}$ & C$_{2}$H$_{4}$ & Broken: C--H: 1.110 to 3.471~\AA. \\
\makecell[tl]{\texttt{transfer\_id\_183\_}\\\texttt{5475\_2\_000-2\_neb1.0}} & \makecell[tl]{H transfer\\(000), site~2} & Rh$_{3}$S$_{2}$Sn$_{2}$ & C$_{2}$H$_{2}$ & Broken: C--H: 1.104 to 3.221~\AA. Formed: C--H: 2.932 to 1.103~\AA. \\
\makecell[tl]{\texttt{transfer\_id\_196\_}\\\texttt{3039\_21\_111-0\_neb1.0}} & \makecell[tl]{H transfer\\(111), site~0} & RhTc$_{3}$ & C$_{2}$H & Broken: C--C: 1.411 to 4.641~\AA. Formed: C--H: 3.058 to 1.108~\AA. \\
\makecell[tl]{\texttt{transfer\_id\_202\_}\\\texttt{9291\_12\_211-1\_neb1.0}} & \makecell[tl]{H transfer\\(211), site~1} & AlAu$_{2}$Ti & CH$_{2}$NO & Broken: C--H: 1.102 to 3.376~\AA. Formed: H--N: 3.099 to 1.035~\AA. \\
\makecell[tl]{\texttt{transfer\_id\_302\_}\\\texttt{8753\_3\_000-3\_neb1.0}} & \makecell[tl]{H transfer\\(000), site~3} & Fe$_{3}$MnZr$_{2}$ & C$_{2}$H$_{4}$ & Broken: C--H: 1.106 to 4.249~\AA. Formed: C--H: 3.098 to 1.108~\AA. \\
\makecell[tl]{\texttt{transfer\_id\_339\_}\\\texttt{7593\_0\_211-0\_neb1.0}} & \makecell[tl]{H transfer\\(211), site~0} & GaOsSc$_{2}$ & CH$_{3}$O & Broken: H--O: 0.981 to 3.100~\AA. Formed: C--H: 4.227 to 1.112~\AA. \\
\makecell[tl]{\texttt{transfer\_id\_349\_}\\\texttt{8583\_22\_011-2\_neb1.0}} & \makecell[tl]{H transfer\\(011), site~2} & CrSe$_{4}$Ti$_{2}$ & C$_{2}$H$_{4}$ & Broken: C--H: 1.100 to 5.367~\AA. Formed: C--H: 3.907 to 1.100~\AA. \\
\makecell[tl]{\texttt{transfer\_id\_351\_}\\\texttt{2818\_1\_000-2\_neb1.0}} & \makecell[tl]{H transfer\\(000), site~2} & PtY$_{3}$ & CH$_{2}$O & Broken: H--O: 0.977 to 2.867~\AA. Formed: C--H: 3.536 to 1.116~\AA. \\
\makecell[tl]{\texttt{transfer\_id\_452\_}\\\texttt{6217\_1\_111-1\_neb1.0}} & \makecell[tl]{H transfer\\(111), site~1} & Ga$_{4}$ScV$_{2}$ & CH$_{2}$O & Broken: H--O: 0.966 to 3.056~\AA. Formed: C--H: 4.003 to 1.118~\AA. \\
\makecell[tl]{\texttt{transfer\_id\_477\_}\\\texttt{3415\_21\_122-1\_neb1.0}} & \makecell[tl]{H transfer\\(122), site~1} & PPd$_{4}$ & C$_{2}$H & Broken: C--C: 1.319 to 4.071~\AA. Formed: C--H: 2.841 to 1.102~\AA. \\
\makecell[tl]{\texttt{transfer\_id\_510\_}\\\texttt{1150\_20\_111-9\_neb1.0}} & \makecell[tl]{H transfer\\(111), site~9} & Pb$_{5}$Rh$_{4}$ & CH$_{4}$O & Broken: C--O: 1.417 to 4.174~\AA. Formed: H--O: 2.753 to 0.977~\AA. \\
\makecell[tl]{\texttt{transfer\_id\_538\_}\\\texttt{11344\_0\_211-1\_neb1.0}} & \makecell[tl]{H transfer\\(211), site~1} & Ir$_{2}$Nb$_{3}$Se$_{10}$ & CH$_{3}$O & Broken: H--O: 0.972 to 2.466~\AA. Formed: C--H: 2.921 to 1.139~\AA. \\
\makecell[tl]{\texttt{transfer\_id\_548\_}\\\texttt{1869\_7\_222-0\_neb1.0}} & \makecell[tl]{H transfer\\(222), site~0} & TiH & CH$_{2}$N & Broken: C--H: 1.115 to 3.339~\AA. Formed: H--N: 3.106 to 1.034~\AA. \\
\makecell[tl]{\texttt{transfer\_id\_632\_}\\\texttt{4044\_2\_211-0\_neb1.0}} & \makecell[tl]{H transfer\\(211), site~0} & AlNi$_{2}$V & C$_{2}$H$_{2}$ & Broken: C--H: 1.148 to 3.940~\AA. Formed: C--H: 4.146 to 1.118~\AA. \\
\makecell[tl]{\texttt{transfer\_id\_671\_}\\\texttt{50\_23\_211-0\_neb1.0}} & \makecell[tl]{H transfer\\(211), site~0} & Ta & C$_{3}$H$_{3}$ & Broken: C--H: 1.171 to 3.004~\AA. Formed: C--H: 3.419 to 1.120~\AA. \\
\makecell[tl]{\texttt{transfer\_ood\_52\_}\\\texttt{6141\_15\_111-3\_neb1.0}} & \makecell[tl]{H transfer\\(111), site~3} & Ga$_{4}$V$_{2}$Zr & C$_{3}$H$_{2}$O & Broken: C--O: 1.410 to 6.547~\AA. Formed: C--O: 4.547 to 1.193~\AA. \\
\makecell[tl]{\texttt{transfer\_ood\_80\_}\\\texttt{4786\_1\_100-1\_neb1.0}} & \makecell[tl]{H transfer\\(100), site~1} & InNTi$_{2}$ & CH$_{2}$O & Broken: H--O: 0.976 to 2.612~\AA. Formed: C--H: 3.315 to 1.126~\AA. \\
\makecell[tl]{\texttt{transfer\_ood\_111\_}\\\texttt{6487\_2\_111-0\_neb1.0}} & \makecell[tl]{H transfer\\(111), site~0} & InY$_{2}$Zn & C$_{2}$H$_{2}$ & Broken: C--H: 1.116 to 3.935~\AA. Formed: C--H: 3.247 to 1.113~\AA. \\
\makecell[tl]{\texttt{transfer\_ood\_160\_}\\\texttt{10666\_22\_122-2\_neb1.0}} & \makecell[tl]{H transfer\\(122), site~2} & Ni$_{3}$S$_{8}$Ta$_{2}$ & C$_{2}$H$_{4}$ & Broken: C--H: 1.110 to 3.951~\AA. Formed: C--H: 3.545 to 1.098~\AA. \\
\makecell[tl]{\texttt{transfer\_ood\_205\_}\\\texttt{9385\_5\_222-0\_neb1.0}} & \makecell[tl]{H transfer\\(222), site~0} & As$_{2}$TiV & C$_{2}$H$_{2}$O & Broken: C--H: 1.119 to 3.963~\AA. Formed: C--H: 4.285 to 1.110~\AA. \\
\makecell[tl]{\texttt{transfer\_ood\_216\_}\\\texttt{10686\_6\_211-0\_neb1.0}} & \makecell[tl]{H transfer\\(211), site~0} & CaNi$_{2}$P$_{2}$ & C$_{2}$H$_{2}$O & Broken: H--O: 0.983 to 3.875~\AA. Formed: C--H: 3.092 to 1.132~\AA. \\
\makecell[tl]{\texttt{transfer\_ood\_439\_}\\\texttt{945\_12\_211-1\_neb1.0}} & \makecell[tl]{H transfer\\(211), site~1} & Pb$_{2}$Pt & CH$_{2}$NO & Broken: C--H: 1.102 to 4.233~\AA. Formed: H--N: 3.101 to 1.049~\AA. \\
\makecell[tl]{\texttt{transfer\_ood\_444\_}\\\texttt{6233\_12\_211-0\_neb1.0}} & \makecell[tl]{H transfer\\(211), site~0} & AuScSn & CH$_{2}$NO & Broken: C--H: 1.109 to 4.379~\AA. Formed: H--N: 3.963 to 1.038~\AA. \\
\makecell[tl]{\texttt{transfer\_ood\_484\_}\\\texttt{370\_3\_111-1\_neb1.0}} & \makecell[tl]{H transfer\\(111), site~1} & CoGa$_{3}$ & C$_{2}$H$_{4}$ & Broken: C--H: 1.110 to 3.751~\AA. \\
\makecell[tl]{\texttt{transfer\_ood\_499\_}\\\texttt{8229\_23\_100-3\_neb1.0}} & \makecell[tl]{H transfer\\(100), site~3} & Al$_{3}$ZnZr$_{2}$ & C$_{3}$H$_{3}$ & Broken: C--H: 1.160 to 3.206~\AA. Formed: C--H: 3.307 to 1.119~\AA. \\
\makecell[tl]{\texttt{transfer\_ood\_583\_}\\\texttt{9984\_1\_211-0\_neb1.0}} & \makecell[tl]{H transfer\\(211), site~0} & NbNiP$_{2}$ & CH$_{2}$O & Broken: H--O: 0.976 to 2.992~\AA. Formed: C--H: 2.981 to 1.098~\AA. \\
\makecell[tl]{\texttt{transfer\_ood\_592\_}\\\texttt{8220\_5\_100-9\_neb1.0}} & \makecell[tl]{H transfer\\(100), site~9} & FeHf$_{9}$Mo$_{4}$ & C$_{2}$H$_{2}$O & Broken: C--H: 1.125 to 3.566~\AA. Formed: C--H: 3.268 to 1.119~\AA. \\
\makecell[tl]{\texttt{transfer\_ood\_615\_}\\\texttt{9965\_23\_111-3\_neb1.0}} & \makecell[tl]{H transfer\\(111), site~3} & GeZn$_{2}$Zr & C$_{3}$H$_{3}$ & Broken: C--H: 1.104 to 5.435~\AA. Formed: C--H: 4.737 to 1.116~\AA. \\
\makecell[tl]{\texttt{transfer\_ood\_641\_}\\\texttt{11049\_9\_211-0\_neb1.0}} & \makecell[tl]{H transfer\\(211), site~0} & Au$_{2}$CsRb & CH$_{3}$N & Broken: C--H: 1.113 to 4.586~\AA. Formed: H--N: 4.551 to 1.035~\AA. \\
\makecell[tl]{\texttt{transfer\_ood\_697\_}\\\texttt{11056\_19\_111-2\_neb1.0}} & \makecell[tl]{H transfer\\(111), site~2} & IrSbTe & H$_{2}$N$_{3}$ & Broken: N--N: 1.274 to 4.870~\AA. Formed: N--N: 3.374 to 1.215~\AA. \\

\end{longtable}
\normalsize  

\setlength{\tabcolsep}{\savedTabColSep}%
\renewcommand{\arraystretch}{\savedArrayStretch}%

\subsection{Failure mode analysis for OC20NEB Cases}
\label{failure-analysis}

\begin{figure}[t]
  \centering 
  \includegraphics[width=0.6\linewidth]{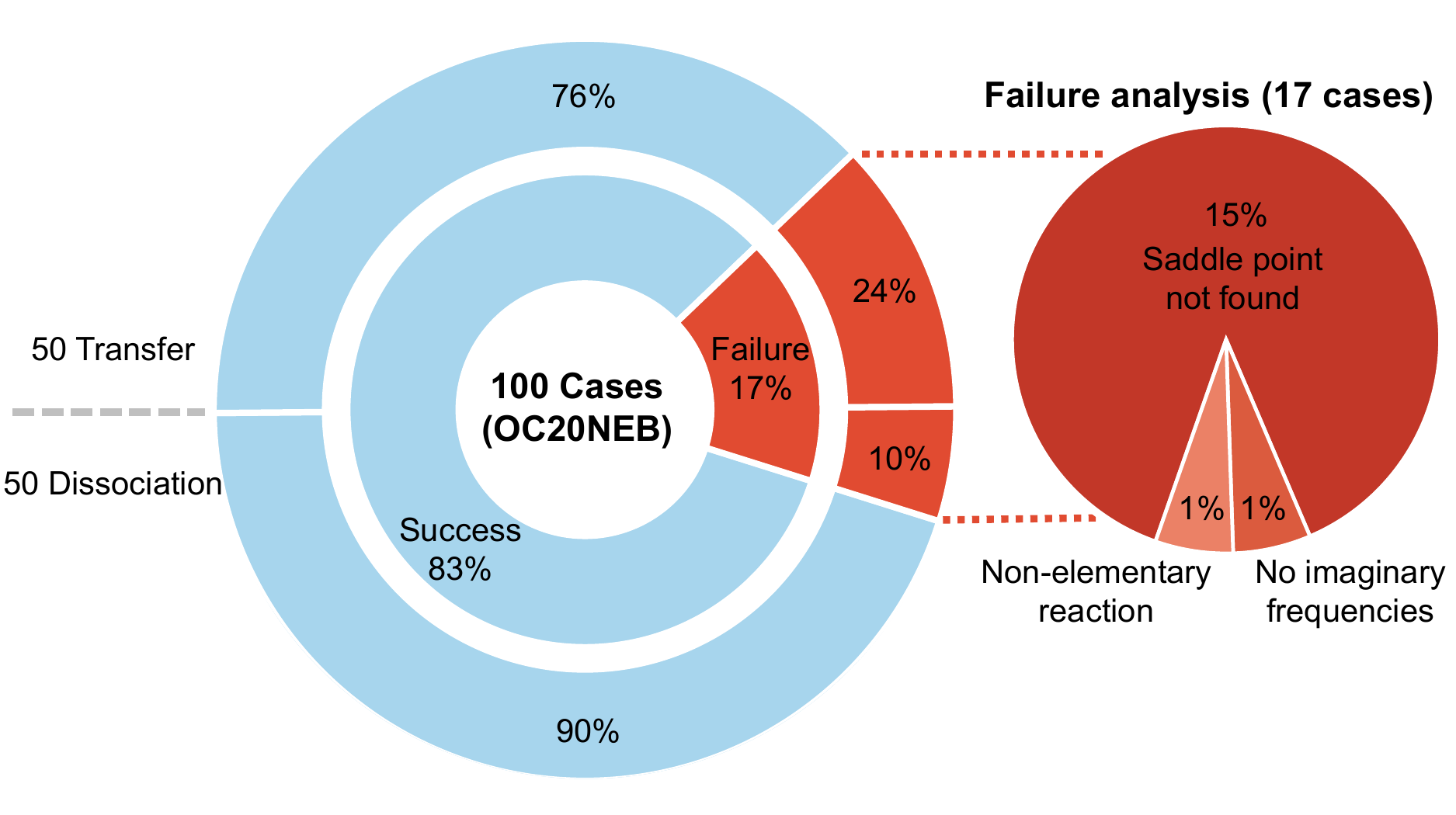}
    \caption{TSAgent performance on a diverse subset of the OC20NEB heterogeneous catalysis transition states benchmark. \textit{Left} Overall success/failure split and per-class success rates. \textit{Right} Breakdown of the 17 failure modes.}
  \label{si:fig:oc20neb_results}
\end{figure}

A TS search is marked as a failure if the agent cannot locate the saddle point corresponding to the target TS in under 5 replans (Figure~\ref{si:fig:oc20neb_results}). For 15\% of total cases, the agent fails to locate a saddle point because it is unable to converge the NEB pathway to sufficiently low inter-atomic forces. In other cases, either no valid imaginary frequency is confirmed even though a saddle point is found, or the agent ran out of compute budget decomposing a multi-step reaction. 

\subsection{UMA Baseline Implementation}
\label{app:uma_baseline}

The UMA baseline provides a point of comparison for TSAgent that reflects the current state of the art in MLIP-accelerated NEB, evaluated at the same level of task difficulty and on the same reaction set as our DFT-level results. Our implementation follows the methodology of CatTSunami~\cite{ocp-neb}, which demonstrated that graph neural network potentials pretrained on the OC20 dataset can perform zero-shot NEB searches on heterogeneous catalyst surfaces at a fraction of the cost of DFT. The key departure from the original work is the choice of potential: CatTSunami evaluated EquiformerV2~\cite{equiformer-v2} as its top-performing model; we replace it with UMA~\cite{uma} (\texttt{uma-s-1p2}), a more recent foundation potential from Meta FAIR that is trained on half a billion unique 3D atomic structures spanning molecules, materials, and catalysts. UMA is accessed via the \texttt{FAIRChemCalculator} interface from the \texttt{fairchem} library with \texttt{task\_name=``oc20''} to apply the OC20-specific energy reference and output head, matching the evaluation protocol of the benchmark.

\paragraph{NEB procedure.}
For each reaction, we read the ten interpolated images stored in the OC20NEB trajectory file as the initial band, consistent with the CatTSunami protocol of using the pre-generated interpolated endpoints rather than DFT-relaxed geometries. Each image is assigned an independent UMA calculator instance. The band is optimized using ASE's \texttt{DyNEB} (dynamic NEB), which skips gradient evaluations on images that have already converged locally, reducing unnecessary force calls in flat regions of the path. We use a fixed spring constant of $k = 1.0$~eV/\AA$^2$ and a BFGS outer optimizer throughout. The optimization proceeds in two stages. In the first stage, the climbing-image modification is disabled and the band is relaxed until the maximum per-image force falls below 0.45~eV/\AA\ or 200 BFGS steps are exhausted. The climbing-image variant is activated only if the first stage converges to this threshold, consistent with the rationale that a poorly resolved band provides an unreliable starting point for the climbing-image constraint. When activated, the second stage is allocated an independent budget of 300 steps and optimized to a tighter convergence threshold of 0.05~eV/\AA. A NEB run is considered converged if and only if the second stage reaches the target force criterion; runs where the first stage fails to converge or the second stage exhausts its step budget are marked as unconverged. The highest-energy internal image at the end of optimization is taken as the candidate TS geometry, and the forward barrier is computed as the energy difference between that image and the initial state.

\paragraph{Vibrational analysis.}
To apply the same success criterion as TSAgent and the human experts, we follow the converged NEB with a finite-difference vibrational frequency analysis on the candidate TS image. Frequencies are computed with ASE's \texttt{Vibrations} module using a two-point finite difference scheme, restricting displacements to the subset of atoms that are not held fixed by the slab constraints. An imaginary frequency is counted as physically significant if its magnitude exceeds 10~meV ($\approx 80$~cm$^{-1}$). A case is marked as a success if the NEB converges, the candidate image carries exactly one significant imaginary frequency, and the forward barrier is positive. The UMA baseline thus applies the same three-part validation gate as TSAgent, ensuring that the comparison is not confounded by differences in the success criterion.

\section{Benchmarking TSAgent Against Human Experts: Experimental Details, Evaluation Metrics, Dataset, and Standard Operating Procedure}
\label{si:sec:human-expert-benchmark}

This appendix documents the full Standard Operating Procedure (SOP) used for the human-expert benchmark reported in Section~\ref{sec:human-expert-benchmark} of the main text. It defines the experiment design and metrics, the fixed compute environment, the per-step DFT settings for geometry optimization, nudged elastic band, and vibrational frequency analysis, and the ten benchmark cases. The same protocol is binding on both TSAgent and the three human experts (HE01--HE03).

\subsection{Experiment Design and Evaluation Metrics}
\label{si:sec:experiment-design}

\paragraph{Controlled common-settings design.}
The benchmark uses a controlled common-settings design: all final simulated results must use the same fixed physical model and execution environment. All calculations are run on Pittsburgh Supercomputing Center machine - Bridges2 (RM queue, one node, 128 cores by default) using VASP~6.3+vtst-intel. A budget of 20{,}000 core-hours per case is enforced for every operator; attempts that exhaust this budget without producing a valid TS are labeled \emph{resource limited}.

Operator choice variables include: initialization decisions; DFT step and convergence parameter choices beyond the minimum common requirements defined in the SOP; handling of failed relaxations; NEB image count and spring-constant selection; restart strategy; diagnostic interpretation; and stopping decisions prior to budget exhaustion. The comparison is therefore not a test of different density exchange-correlation functionals, energy cutoffs, or compute queues, but of \emph{search policy and debugging behavior} under a shared computational protocol.

Any setting explicitly defined in the templates must be identical between the agent and the human experts in their final reported results. Any tag intentionally left as an operator-choice (``free'') variable is part of the comparison. Every attempt is fully logged, including the GO/NEB/VFA settings used, failures and diagnostics, what was changed, and the rationale for each change.

\paragraph{Agent and human operator procedures.}
For TSAgent, all decisions including failure diagnosis and replanning are produced autonomously by the workflow with no mid-run human intervention beyond passive inspection of job status. TSAgent uses the same simulation configuration as described in Section~\ref{oc20neb_tsa_config}. For human experts, the same input cases and compute limits are provided, but each expert is free to apply their own practical strategy within the SOP constraints.
Human experts are not required to mimic TSAgent's internal policy. This design reflects the intended comparison: an autonomous agent versus skilled manual operation under a shared protocol.

\paragraph{Success criterion.}
Each case follows the same three-step order: geometry optimization (GO), nudged elastic band (NEB), and vibrational frequency analysis (VFA). All attempts are logged, including the GO/NEB/VFA settings used, diagnostics, failure labels, parameter changes, and the rationale for each intervention. A case is counted as solved only when: 1) GO and the saddle-point calculation both converge at the minimum settings defined in the SOP; and 2) VFA confirms exactly one imaginary frequency with magnitude ${>}10$~meV (80.65~cm$^{-1}$).

Cases that do not satisfy both criteria---for example, those with zero imaginary frequencies, more than one imaginary frequency, or an imaginary frequency below the threshold---are not counted as successful transition states. This criterion prevents ambiguous near-saddle structures, non-converged endpoints, and spurious NEB candidates from being counted as successes.

\paragraph{Operator efforts.}
Core-hours measure resource consumption but do not capture the manual burden of inspecting structures, diagnosing failed optimizations, modifying DFT input files, resubmitting jobs, and deciding whether a saddle-point candidate warrants VFA. Even when two operators consume comparable core-hours, the supervision burden may differ substantially.
For each human expert, operator efforts is logged only for direct work on the benchmark, including setup, job review, diagnosis, parameter selection, and documentation.
We treat operator effort as a separate axis from compute cost (Table~\ref{tab:human_aggregate_results} in the main text).

\paragraph{Reported metrics.}
Three metrics are reported for the comparison: (i) TS success rate, defined as the percentage of benchmark cases that satisfy the success criterion; (ii) average core-hours per successful case, capturing compute consumption on the cases each operator solved; and (iii) operator effort, measured in active minutes per successful case as defined above.

\subsection{DFT Settings}
\label{si:sec:dft-settings-HEeval}

A common set of VASP parameters is fixed across all GO, NEB, and VFA steps for every case. KPOINTS sampling uses a Monkhorst--Pack mesh of 3$\times$3$\times$1. The plane-wave basis and electronic-structure settings are \texttt{ALGO}=Normal, \texttt{PREC}=Normal, \texttt{ENCUT}=450~eV, \texttt{GGA}=RP, \texttt{IVDW}=12, \texttt{ISMEAR}=0, \texttt{SIGMA}=0.05, \texttt{ISPIN}=1, and \texttt{ISYM}=0. Ionic-relaxation settings are \texttt{ISIF}=2 and \texttt{EDIFFG}=$-5.00\times10^{-2}$~eV/\AA. Dipole corrections are enabled along the surface normal with \texttt{IDIPOL}=3 and \texttt{LDIPOL}=.TRUE.; and projection is performed in real space (\texttt{LREAL}=Auto). Parallelization uses \texttt{NCORE}=8 by default; this is overridden in NEB (see below). The two consistency tags \texttt{DIPOL} and \texttt{LMAXMIX} are not fixed to a particular value, but whichever values an operator selects for a case must be reused identically across that case's GO, NEB, and VFA steps.

\paragraph{Geometry optimization (GO) settings.}
\label{si:sec:go-settings}
GO inherits the common settings above with no overrides. The operator-choice (free) tags are \texttt{POTIM}, \texttt{IBRION}, and \texttt{NSW}; each operator selects values per case and logs the rationale.

\paragraph{Nudged elastic band (NEB) settings.}
\label{si:sec:neb-settings}
NEB inherits the common settings and additionally converges the final calculations with \texttt{LCLIMB}=.TRUE. The operator-choice tags are \texttt{POTIM}, \texttt{IBRION}, \texttt{NSW}, \texttt{NCORE}, \texttt{SPRING}, \texttt{MAXMOVE}, \texttt{IMAGES}, and \texttt{IOPT}. Because NEB parallelizes across images, \texttt{NCORE} must be coordinated with the chosen \texttt{IMAGES} on a 128-core Bridges2 RM node so that the total MPI task count is at or near 128. Recommended \texttt{IMAGES}/\texttt{NCORE}/total-tasks pairings are 5/5/125, 6/7/126, 7/6/126, 8/8/128, 9/7/126, and 10/6/120.

\paragraph{Vibrational frequency analysis (VFA) settings.}
\label{si:sec:vfa-settings}
VFA inherits the common settings and additionally fixes a tighter electronic convergence \texttt{EDIFF}=$1\times10^{-6}$~eV. It operates in finite-difference mode with \texttt{IBRION}=5, \texttt{NFREE}=2, and \texttt{NSW}=1 on the converged saddle-point geometry produced by NEB. \texttt{POTIM} is the only operator-choice tag.

\subsection{Benchmark Cases}
\label{si:sec:benchmark-cases}


\begingroup
\setlength{\tabcolsep}{4pt}
\renewcommand{\arraystretch}{1.12}
\begin{table}[ht]
\centering
\scriptsize
\caption{%
    Benchmark cases and chemistry for the human expert evaluation.
    The benchmark spans both dissociation and H-transfer reactions
    across diverse OC20NEB surfaces.
}
\label{si:tab:human_benchmark_cases}
\begin{tabularx}{\linewidth}{%
    >{\raggedright\arraybackslash}p{0.22\linewidth}
    >{\raggedright\arraybackslash}p{0.15\linewidth}
    >{\raggedright\arraybackslash}p{0.19\linewidth}
    >{\raggedright\arraybackslash}p{0.08\linewidth}
    >{\raggedright\arraybackslash}X}
\toprule
\textbf{Case}
    & \textbf{Split / facet}
    & \textbf{Surface}
    & \textbf{Ads.}
    & \textbf{Reaction and bond change} \\
\midrule

\makecell[tl]{\texttt{dissociation\_id\_250\_}\\
              \texttt{4710\_31\_111-0\_neb1.0}}
    & \makecell[tl]{dissociation \\ (111), site~0}
    & \makecell[tl]{PtSiTi \\ ternary intermetallic, \\ silicide-like}
    & C$_{2}$H$_{2}$
    & C$_{2}$H$_{2}$ splits into two CH fragments;
      C--C broken at 1.454~\AA. \\[4pt]

\makecell[tl]{\texttt{dissociation\_id\_502\_}\\
              \texttt{9246\_48\_211-0\_neb1.0}}
    & \makecell[tl]{dissociation \\ (211), site~0}
    & \makecell[tl]{Al$_{2}$TiZn \\ ternary alloy, \\ intermetallic}
    & CN
    & CN dissociates into separate C and N; C--N broken at 1.309~\AA. \\[4pt]

\makecell[tl]{\texttt{dissociation\_ood\_346\_}\\
              \texttt{1613\_12\_211-0\_neb1.0}}
    & \makecell[tl]{dissociation \\ (211), site~0}
    & \makecell[tl]{CdPd$_{3}$ \\ bimetallic alloy, \\ intermetallic}
    & CH$_{2}$O
    & C--H cleavage in a CHOH-like adsorbate; H relocates near Pd. \\[4pt]

\makecell[tl]{\texttt{dissociation\_ood\_417\_}\\
              \texttt{4057\_50\_222-1\_neb1.0}}
    & \makecell[tl]{dissociation \\ (222), site~1}
    & \makecell[tl]{AlFeRh$_{2}$ \\ ternary alloy, \\ intermetallic}
    & H$_{2}$N
    & NH$_{2}$-like dissociation to NH$+$H;
      N--H broken at 1.028~\AA. \\[4pt]

\makecell[tl]{\texttt{dissociation\_ood\_684\_}\\
              \texttt{2600\_11\_111-1\_neb1.0}}
    & \makecell[tl]{dissociation \\ (111), site~1}
    & \makecell[tl]{Al$_{3}$Zr \\ binary alloy, \\ intermetallic}
    & CH$_{2}$O
    & O--H cleavage in a CHOH-like adsorbate;
      C--H and C--O remain intact. \\

\midrule

\makecell[tl]{\texttt{transfer\_id\_246\_}\\
              \texttt{9298\_5\_211-4\_neb1.0}}
    & \makecell[tl]{H transfer \\ (211), site~4}
    & \makecell[tl]{CrFeGe$_{2}$ \\ ternary germanide, \\ intermetallic}
    & C$_{2}$H$_{2}$O
    & H transfers between C fragments;
      C--O shortens from 1.386 to 1.181~\AA. \\[4pt]

\makecell[tl]{\texttt{transfer\_id\_367\_}\\
              \texttt{7714\_8\_122-5\_neb1.0}}
    & \makecell[tl]{H transfer \\ (122), site~5}
    & \makecell[tl]{Cu$_{3}$PS$_{4}$ \\ copper thiophosphate, \\ sulfide}
    & C$_{3}$H$_{6}$O
    & H transfers to a separate C fragment;
      C--O shortens from 1.473 to 1.227~\AA. \\[4pt]

\makecell[tl]{\texttt{transfer\_id\_472\_}\\
              \texttt{786\_1\_100-2\_neb1.0}}
    & \makecell[tl]{H transfer \\ (100), site~2}
    & \makecell[tl]{CrS$_{2}$ \\ transition-metal \\ sulfide}
    & CH$_{2}$O
    & H transfers from O to C; O--H breaks and a new C--H forms. \\[4pt]

\makecell[tl]{\texttt{transfer\_ood\_372\_}\\
              \texttt{3585\_9\_111-0\_neb1.0}}
    & \makecell[tl]{H transfer \\ (111), site~0}
    & \makecell[tl]{CdPd$_{3}$ \\ bimetallic alloy, \\ intermetallic}
    & CH$_{3}$N
    & H transfers from C to N; C--H breaks and N--H forms. \\[4pt]

\makecell[tl]{\texttt{transfer\_ood\_429\_}\\
              \texttt{9454\_0\_222-2\_neb1.0}}
    & \makecell[tl]{H transfer \\ (222), site~2}
    & \makecell[tl]{H$_{4}$NbV \\ mixed Nb--V \\ hydride}
    & CH$_{3}$O
    & H transfers from O to C; O--H breaks and a new C--H forms. \\

\bottomrule
\end{tabularx}
\end{table}
\endgroup  

\subsection{Transition State Energies for Ten OC20NEB Cases}
\label{si:sec:HEeval_TSEnergies}

\begingroup
\setlength{\tabcolsep}{4pt}
\renewcommand{\arraystretch}{1.12}
\begin{table}[ht]
\centering
\scriptsize
\caption{%
    Transition state energies for ten OC20NEB benchmark cases.
    Energies are reported as $\Delta E_{\mathrm{TS-IS}}$ (eV).}
\label{si:tab:transition_state_energies}
\begin{tabularx}{\linewidth}{%
    >{\raggedright\arraybackslash}p{0.48\linewidth}
    *{4}{>{\centering\arraybackslash}X}}
\toprule
\textbf{Reaction system ID}
    & \textbf{HE01} & \textbf{HE02} & \textbf{HE03}
    & \textbf{TSAgent} \\
\midrule
dissociation\_id\_250\_4710\_31\_111-0      & --   & --   & --   & --   \\
dissociation\_id\_502\_9246\_48\_211-0      & 2.04 & 2.01 & 2.00 & 2.01 \\
dissociation\_ood\_346\_1613\_12\_211-0     & 0.97 & 0.35 & 0.97 & --   \\
dissociation\_ood\_417\_4057\_50\_222-1     & 1.09 & 0.95 & 0.95 & --   \\
dissociation\_ood\_684\_2600\_11\_111-1     & 0.85 & 0.85 & 0.85 & 0.84 \\
transfer\_id\_246\_9298\_5\_211-4           & --   & --   & --   & 0.12 \\
transfer\_id\_367\_7714\_8\_122-5           & 1.81 & 1.81 & 1.81 & 1.83 \\
transfer\_id\_472\_786\_1\_100-2            & 0.77 & 1.37 & 0.76 & 0.91 \\
transfer\_ood\_372\_3585\_9\_111-0          & 0.94 & 0.35 & --   & 0.61 \\
transfer\_ood\_429\_9454\_0\_222-2          & 0.45 & 0.45 & --   & 0.59 \\
\bottomrule
\end{tabularx}
\end{table}
\endgroup

\paragraph{Discussion.}
Variability in TS energies across/within human experts and TSAgent is expected on theoretical grounds and is not computational error. NEB is a local optimizer \cite{henkelman-ci-neb}, so cropping or re-segmenting the band, the interpolation scheme, image count, and the climbing-image force tolerance each bias which saddle is reached \cite{smidstrup_improved_2014, zarkevich_nudged-elastic_2015}. Rigorously, the unbiased object is the transition-path ensemble between defined reactant and product basins. In practice one retains the lowest-energy first-order saddle and runs several searches per step \cite{dellago_TSpath_2002}. Our data in Table~\ref{si:tab:transition_state_energies} show both regimes, barriers for \texttt{transfer\_id\_4722} span 0.76--1.37~eV across HE01--03, and the HE02 (1.34~eV) vs TSAgent (0.91~eV) geometries are distinct in energy. For \texttt{transfer\_id\_246} every HE failed, whereas the TSAgent, by iteratively intervening on its single initialization until convergence, recovered a saddle at 0.12~eV and found a TS.

\section*{Licenses for Existing Assets}
\label{app:licenses}

The OC20NEB dataset is released by Meta AI as part of the Open Catalyst Project
under a Creative Commons Attribution 4.0 International (CC BY 4.0) license,
consistent with the broader OC20 data release.
VASP 6.3, including the VTST extensions for nudged elastic band calculations,
is commercial software accessed under an institutional academic license.
GPT-5.4 is accessed via the OpenAI API under OpenAI's standard services agreement.
The Atomic Simulation Environment (ASE) is released under the GNU Lesser General
Public License v2.1 (LGPL-2.1); the remainder of the Python software stack
(NumPy, SciPy, matplotlib, and related packages) is distributed under permissive
MIT or BSD licenses.





\end{document}